\documentclass[11pt]{article}
\usepackage{amsmath}
\usepackage{amsfonts}
\usepackage{latexsym}

\newtheorem{theorem}{Theorem}
\newtheorem{lemma}{Lemma}
\newtheorem{corollary}{Corollary}
\newtheorem{proposition}{Proposition}
\newtheorem{definition}{Definition}
\newtheorem{remark}{{\bf Remark}}
\newlength{\dinwidth}
\newlength{\dinmargin}
\def\beqs{\begin{displaymath}}
\def\eeqs{\end{displaymath}}
\def\beqn{\begin{eqnarray}}
\def\eeqn{\end{eqnarray}}

\def\Lie{{\rm Lie}}

\def\l{\lambda}
\def\I{{\rm I}}
\def\C{\mathbb {C}}
\def\Z{\mathbb {Z}}
\def\R{\mathbb {R}}
\def\N{\mathbb{N}}
\def\F{{\cal F}}

\def\a{\alpha}
\def\b{\beta}
\def\g{\gamma}
\def\d{\partial}
\def\hsp{\hspace{0.5cm}}
\def\L{{\cal L}}

\def\imb{\mathrm{Im}\mathbb B}
\def\B{\mathbb B}
\def\res{\mathrm{res}}
\def\Qbar{\bar{Q}}
\def\Pbar{\bar{P}}
\def\lb{\bar{\l}}

\def\ib{{\bar{i}}}
\def\kb{{\bar{k}}}
\def\nb{{\bar{n}}}
\def\zb{{\bar{z}}}
\def\vp{\mathrm{v.p.}}
\def\jb{{\bar{j}}}

\def \q {{\bf {q}}}
\def\Wq{W_\q}

\def\iQ{{\scriptscriptstyle{Q}}}
\def\iA{{\scriptscriptstyle{A}}}
\def\iB{{\scriptscriptstyle{B}}}
\def\iC{{\scriptscriptstyle{C}}}

\def\iP{{\scriptscriptstyle{P}}}
\def\iI{{\scriptscriptstyle{{\rm I}}}}
\def\iW{{\scriptscriptstyle{W}}}

\def\iL{{\scriptscriptstyle{L}}}
\def\iE{{\scriptscriptstyle{E}}}
\def\iF{{\scriptscriptstyle{F}}}
\def\iT{{\scriptscriptstyle{T}}}
\def\iPhi{{\scriptscriptstyle{\Phi}}}
\def\iWq{{\scriptscriptstyle{\Wq}}}
\def\iOmega{{\scriptscriptstyle{\Omega}}}
\def\iOmt{{\scriptscriptstyle{\Omt}}}
\def\h{{\scriptscriptstyle{(1,0)}}}
\def\ah{{\scriptscriptstyle{(0,1)}}}

\def \surf {\L}

\def\covM{\widehat{M}}
\def\z{\varsigma}

\def\dim{L}

\def \Bt {{\B^\iOmega}}
\def \Omt {\Omega_\q}
\def \Brt {B_\q}
\def \Div {{\cal D}_\q}
\def \DivR {{\cal D}^\iOmega_\q}

\makeatletter
\@addtoreset{equation}{section}
\makeatother

\begin{document}

\title{Deformations of Frobenius structures on Hurwitz spaces}      
\author{Vasilisa Shramchenko}        
\date{}   
\maketitle
\begin{center}
Department of Mathematics and Statistics, Concordia University\\
7141 Sherbrooke West, Montr\'eal H4B 1R6, Qu\'ebec, Canada
\\e-mail: vasilisa@mathstat.concordia.ca
\end{center}

\vspace{0.8cm}
\textbf{Abstract.} Deformations of Dubrovin's Hurwitz Frobenius manifolds are constructed. The deformations depend on $g(g+1)/2$ complex parameters where $g$ is the genus of the corresponding Riemann surface. In genus one, the flat metric of the deformed Frobenius manifold coincides with a metric associated with a one-parameter family of solutions to the Painlev\'e-VI equation with coefficients $(1/8,-1/8,1/8,3/8)\;.$ Analogous deformations of real doubles of the Hurwitz Frobenius manifolds are also found; these deformations depend on $g(g+1)/2$ real parameters. 
\vspace{0.2cm}

\tableofcontents

\section{Introduction}

The structure of a Frobenius manifold was introduced in \cite{Dubrovin} (see also \cite{Manin}) to give a geometric reformulation of the Witten-Dijkgraaf-Verlinde-Verlinde (WDVV) system of differential equations on the function $F$ of $n$ variables (\cite{DVV,Witten}):
\beqn 
F_i F_1^{-1} F_j = F_j F_1^{-1} F_i \;, \hsp i,j=1, \dots, n \;,
\label{WDVV}
\eeqn
where $F_i$ is the matrix 
\beqn
(F_i)_{mn} = \frac{\d^3F}{\d t^i\d t^m\d t^n} \;,
\label{matrices}
\eeqn
and 
the function $F$  is such that $F_1$ is a constant nondegenerate matrix, and  there exist constants $\nu_1,\dots,\nu_n,\nu_\iF$ such that for any nonzero constant $\kappa$ the following relation (quasihomogeneity) holds:
\beqn
F( \kappa^{\nu_1}t^1, \dots, \kappa^{\nu_n}t^n ) = \kappa^{\nu_\iF} F( t^1, \dots, t^n) + \mbox{quadratic terms} \;.
\label{quasihomogeneity}
\eeqn
The function $F$ is called the {\it prepotential} of the corresponding Frobenius manifold. 

Here we consider the so-called semisimple Frobenius structures on Hurwitz spaces (a Frobenius manifold is called {\it semisimple} if the associated algebra in the tangent space does not have nilpotents). The Hurwitz space is the space of pairs $(\surf\,,\l)$ modulo an equivalence relation (see Section \ref{Hurwitz}) where $\surf$ is a Riemann surface of genus $g$ and $\l$ is a function on the surface, $\l:\surf \to \C P^1,$ of a fixed degree. The finite critical values of the function $\l$ (semisimplicity implies they are all simple) serve as local coordinates on the Hurwitz space. Frobenius structures on Hurwitz spaces in any genus were originally found in \cite{Dubrovin}. Local coordinates on the Hurwitz space become  canonical coordinates on the Frobenius manifold. In \cite{Dubrovin}, App. I, it is shown that any Frobenius manifold, under some genericity assumption, can be locally described in terms of Hurwitz spaces: for any Frobenius manifold there exists a function of one complex variable (called the {\it superpotential}) meromorphic in some domain in $\C$ and such that canonical coordinates on the Frobenius manifold are given by critical values of this function. If the superpotential can be analytically continued to a meromorphic function on a compact Riemann surface then the corresponding Frobenius manifold is isomorphic to a Hurwitz Frobenius manifold; in this case the Hurwitz space is the space of coverings defined by the superpotential. 
Therefore, one might expect that any natural result concerning Hurwitz Frobenius manifods can be extended to an arbitrary Frobenius manifold. 
In \cite {doubles} new semisimple Frobenius structures which can be considered as  {\it real doubles} of the semisimple Hurwitz Frobenius manifolds of Dubrovin \cite{Dubrovin} were found. Those Frobenius structures are built on Hurwitz spaces considered as real manifolds.

For the simplest Hurwitz space in genus one the Frobenius structure of \cite{Dubrovin} gives the following solution to the WDVV system:
\beqn
F = -\frac{1}{4} t_1 t_2^2 + \frac{1}{2}t_1^2t_3 - \frac{\pi i}{32} t_2^4 \,\gamma\left( 2\pi i t_3 \right) \;,
\label{example}
\eeqn
where $\gamma(\mu)=\theta_1^{\prime\prime\prime}/(3\pi i \theta_1^\prime) \,;$ and $\theta_1(z) = - \theta[\frac{1}{2},\frac{1}{2}](z)$ is the odd elliptic Jacobi theta function. The function $\gamma(\mu)$ satisfies the Chazy equation 
\beqn
\gamma^{\prime\prime\prime} = 6 \gamma \gamma^{\prime\prime} - 9 (\gamma^\prime)^2 \;.
\label{Chazy}
\eeqn
It is known (\cite{Dubrovin}, App. C) that the function of the form (\ref{example}) will still satisfy the WDVV system if the function $\g$ in (\ref{example}) is replaced by an arbitrary solution to the Chazy equation (\ref{Chazy}). 
The general solution to the Chazy equation has the form:
\beqn
 f(\mu) = \gamma \left( \frac{a\mu + b}{c\mu + d} \right) \frac{1}{(c\mu + d)^2} - \frac{2c}{c\mu + d} 
\label{transform_Chazy}
\eeqn
where $ \left( \begin{matrix} a&b\\c&d \end{matrix}  \right) \in SL(2,\C)\;.$

In particular, in the case of $SL(2,\C)$-transformations of the form $\left( \begin{matrix} 1&0\\-1/\q&1 \end{matrix}  \right)$ we get the following solution to WDVV equations: 
\begin{equation}
F = -\frac{1}{4} t_1 t_2^2 + \frac{1}{2}t_1^2t_3 - \frac{\pi i}{32} t_2^4 \left( \frac{1}{(1 - 2\pi i t_3/\q)^2}\gamma\left( \frac{2\pi i t_3}{1 - 2\pi i t_3/\q}\right) + \frac{2}{\q(1 - 2\pi i t_3/\q)}  \right) \;,
\label{transform_example}
\end{equation}
This function is obtained from (\ref{example}) by replacing the function $\g(2 \pi i t_3)$ by $f(2 \pi i t_3)$ from (\ref{transform_Chazy}) with $ \left( \begin{matrix} a&b\\c&d \end{matrix}  \right) = \left( \begin{matrix} 1&0\\-1/\q&1 \end{matrix}  \right)\;.$ If $(1/\q) \in \Z$ then the  solutions (\ref{example}) and (\ref{transform_example}) coincide due to the modular invariance of the function $\g\,;$ for $(1/\q) \notin \Z$ we obtain a one-parameter deformation of the solution (\ref{example}).

The main result of this paper is a generalization of this deformation procedure to semisimple Hurwitz Frobenius manifolds in any genus. Namely, we construct a $g(g+1)/2$-parametric deformation of Dubrovin's Frobenius structures \cite{Dubrovin} on Hurwitz spaces. For the simplest Hurwitz space in genus one our deformation coincides with the deformation (\ref{transform_example}) of the prepotential (and corresponding Frobenius manifold) (\ref{example}). 

The idea of the construction is the following. All ingredients of semisimple Hurwitz Frobenius manifolds of Dubrovin can be conveniently described  in terms of the canonical meromorphic bidifferential $W$ on a Riemann surface $\surf\,.$ The bidifferential $W$ is defined as follows. Introduce on $\L$ a canonical basis of cycles $\{a_k;\,b_k\}\;.$ Then $W(P,Q)$ is a symmetric bidifferential which has a second order pole with biresidue $1$ on the diagonal $P\sim Q$ and has vanishing $a$-periods; it can be expressed in terms of the prime form $E(P,Q)$ as follows $W(P,Q):= d_P d_Q \log E(P,Q)\;.$ For a Hurwitz space of coverings $(\surf\,, \l)$ with simple ramification points $\{P_j\}\;,$ the dependence of the bidifferential $W$ on the branch points $\{\l_j\}$ is given by the Rauch variational formulas \cite{KokKor, Rauch}:
\beqn
\frac{\d W(P,Q)}{\d\l_j}=\frac{1}{2}W(P,P_j)W(Q,P_j)\;,
\label{intro:W-variation}
\eeqn
where $W(P,P_j):=\left({W(P,Q)}/{dx_j(Q)}\right)\vert_{Q=P_j}\;.$

The main ingredient of Frobenius structures is a {\it Darboux-Egoroff} metric. A diagonal metric ${\bf ds^2}=\sum_i g_{ii}(d\l_i)^2$ is called a Darboux-Egoroff metric if it is flat (its curvature tensor vanishes) and potential (there exists a function $U$ such that $g_{ii} = \d_{\l_i}U$ holds for any $i$). The Darboux-Egoroff lemma states that a diagonal metric is potential and flat if its  {\it rotation coefficients} $\beta_{ij}$ defined for $i \neq j$ by
$\beta_{ij}=({\d_{\l_j}\sqrt{g_{ii}}})/{\sqrt{g_{jj}}} $
are symmetric, $\beta_{ij}=\beta_{ji},$ and satisfy the system of equations:
\begin{align}
&\d_{\l_k} \beta_{ij} = \beta_{ik} \beta_{kj} \;,\hsp i,j,k\;\; \mbox{are distinct},
\label{intro:rotsys1} \\
&\sum_k \d_{\l_k} \beta_{ij} = 0 \qquad \mbox {for all} \;\;\beta_{ij}\;.
\label{intro:rotsys2}
\end{align}
 For the family of Hurwitz Frobenius manifolds introduced in \cite{Dubrovin}, the rotation coefficients of the corresponding Darboux-Egoroff metrics are given by $\beta_{ij} = W(P_i,P_j)/2\;.$ These rotation coefficients satisfy equations (\ref{intro:rotsys1}) due to the Rauch formulas (\ref{intro:W-variation}). 

In this work we introduce the following deformation of the bidifferential $W\,:$ 
\beqs
\;\Wq(P,Q) := W(P,Q)-2 \pi i \sum_{k,l=1}^g(\B+\q)_{kl}^{-1}\omega_k(P)\omega_l(Q)\;,
\eeqs
 where $g$ is the genus of the Riemann surface; $\omega_l(Q):=\oint_{b_l} W(P,Q)/(2\pi i)$ form the basis of holomorphic differentials normalized by $\oint_{a_k}\omega_l=\delta_{kl}\,;\;$  $\;\B_{kl}:= \oint_{b_k}\omega_l$ is the matrix of $b$-periods; and $\q$ is a symmetric matrix of parameters constant with respect to $\{\l_j\}$ and $\l\;.$ The matrix $\q$ must be chosen  such that the sum $(\B+\q)$ is not degenerate. 

The bidifferential $\Wq(P,Q)$ turns out to satisfy the following variational formulas which look identical to the variational formulas (\ref{intro:W-variation}) for $W\,:$ 
\beqn
\d_{\l_j} \Wq(P,Q)=\frac{1}{2}\,\Wq(P,P_j) \, \Wq(Q,P_j)\;.
\label{intro-var-Wq}
\eeqn
Therefore, the quantities $\Wq(P_i,P_j)/2$ give rotation coefficients of some Darboux-Egoroff metric: the bidifferential $W$ is symmetric, i.e. $W(P,Q)=W(Q,P)\,;$ the variational formulas (\ref{intro-var-Wq}) imply relations (\ref{intro:rotsys1}) for the rotation coefficients; the equations (\ref{intro:rotsys2}) can be  proven analogously to the case of rotation coefficients given by the bidifferential $W\,.$ The variational formulas (\ref{intro-var-Wq}) hold for the points of the Hurwitz space which do not belong to the divisor defined by the equation $\det\,(\B+\q)=0\;.$ The corresponding Darboux-Egoroff metrics are also defined outside this divisor.  Analogously to \cite{Dubrovin} (see also \cite{doubles}) we find a family of Darboux-Egoroff metrics on Hurwitz spaces with rotation coefficients $\Wq(P_i,P_j)/2$ and build corresponding Frobenius structures. In the limit as some  entries of the matrix $\q$ tend to infinity so that all entries of the  matrix $(\B+\q)^{-1}$ tend to zero (in particular this condition holds if all diagonal entries of the matrix $\q$ tend to infinity and nondiagonal entries remain finite) the bidifferential $\Wq$ turns into $W$ and our Frobenius structures coincide with those of \cite{Dubrovin}. 

The second result of the paper is a construction of real doubles \cite{doubles} of the deformed semi-simple Hurwitz Frobenius manifolds. This is done by introducing deformations of the Schiffer and Bergman kernels. The Schiffer and Bergman  kernels  were used in the construction of real doubles in \cite{doubles}; they are defined by the following formulas:
\begin{align*}
\Omega(P,Q) &:= W(P,Q) - \pi\sum_{k,l=1}^g(\imb)_{kl}^{-1}\omega_k(P)\omega_l(Q) \;,
\\
B(P,\Qbar) &:= \pi \sum_{k,l=1}^g(\imb)_{kl}^{-1}\omega_k(P)\overline{\omega_l(Q)} \;,
\end{align*}
respectively. In the case of a genus zero Riemann surface the Schiffer kernel coincides with $W$ and the Bergman kernel vanishes.

The following alternative definitions \cite{Fay92} independent of the choice of a canonical basis of cycles $\{a_k;b_k\}_{k=1}^g$ on the Riemann surface can be given for the two kernels. The Schiffer kernel is the symmetric bidifferential which has a second order pole along the diagonal $P=Q$ and is such that 
$ p.v. {\iint}_\surf \Omega(P,Q) \; \overline {\omega(P)} = 0 $
holds for any holomorphic differential $\omega$ on the surface. 
The Bergman kernel is a regular bidifferential on $\surf$ holomorphic with respect to its first argument and antiholomorphic with respect to the second one which (up to a factor of $2\pi i$) is a kernel of an integral operator acting in the space $L_2^\h(\surf)$ of (1,0)-forms as an orthogonal projector onto the subspace ${\cal H}^\h(\surf)$ of holomorphic (1,0)-forms. In particular, for any holomorphic differential $\omega$ on the surface $\surf$ the following relation holds: 
$\iint_\surf B(P,\bar{Q}) \; \omega(Q) = 2\pi i \; \omega(P) \;.$

In contrast to the bidifferential $W\;,$ which is holomorphic with respect to the moduli coordinates $\{\l_k\}\,,$ the Schiffer and Bergman kernels depend on the complex structure of the Riemann surface through the branch points $\{\l_k\}$ of the covering $(\surf\,,\l)$ {\it and} their complex conjugates $\{\lb_k\}\,.$ Therefore, in \cite{doubles} the Hurwitz space was considered as a real manifold, i.e. a manifold with coordinates $\{\l_k;\lb_k\}\,.$ A family of Darboux-Egoroff metrics on this real space was found; the rotation coefficients of those metrics are given by the Schiffer and Bergman kernels suitably evaluated at ramification points of the covering. The flatness for the metrics is provided by variational formulas for the kernels $\Omega$ and $B\;.$ Some of the Darboux-Egoroff metrics proved to correspond to Frobenius structures on the Hurwitz space with coordinates $\{\l_k;\lb_k\}\,.$ Those Frobenius structures were called the real doubles of Dubrovin's Hurwitz Frobenius structures.

We introduce the following deformations $\Omt(P,Q)$ and $\Brt(P,Q)$ of the Schiffer and Bergman kernels. Consider the holomorphic differentials 
$ v_k(P) :=  \oint_{b_k} \Omega(P,Q)/({2 \pi i})\,.$ The differential $v_k$ is normalized by the condition that all its $a$- and $b$-periods are purely imaginary except the $a_k$-period: ${\rm Re}\{\oint_{a_j}v_k\}=\delta_{jk}/2 \;$ and 
${\rm Re} \{\oint_{b_j}v_k\} = 0 \;$ 
for $\;j,k = 1,\dots, g\;.$
The matrix $\Bt$ of $b$-periods of differentials $v_k$ (which is symmetric and imaginary) is given by:
$ \Bt :=  \bar{\B}(\bar{\B} - \B )^{-1} \B   \;;$ it is the matrix  of pairwise scalar products of differentials $v_k$ in the space $L_2^\h(\surf)\;,$ i.e. $\iint_\surf \overline{v_k(P)}\wedge v_l(P) = \Bt_{kl}\;.$ Then, if a constant matrix $\q$ is such that $\q=\q^\iT\,,$  $\bar{\q} = - \q$ and the matrix $(\Bt+\q)$ is invertible, we can define the deformed Schiffer and Bergman kernels by:
\begin{align}
\begin{split}
\Omt(P,Q) := \Omega(P,Q) - 2\pi i \sum_{k,l=1}^g(\Bt + \q)_{kl}^{-1} v_k(P) v_l(Q) \;,
\\
\Brt(P,\Qbar) := B(P,\Qbar) - 2\pi i \sum_{k,l=1}^g(\Bt + \q)_{kl}^{-1} v_k(P) \overline{v_l(Q)} \;,
\label{deform-kernels}
\end{split}
\end{align}
respectively. 
The integral operator with the deformed kernel $\Brt(P,\bar{Q})/(2\pi i)$ maps the space $L_2^\h(\surf)$ onto the space ${\cal H}^\h(\surf)$ and acts in the space ${\cal H}^\h(\surf)$ of holomorphic differentials as a linear operator which in the basis $\{v_k\}$ is given by the matrix $\q(\Bt+\q)^{-1}\,.$ Similarly, the action of the integral operator with the kernel $\Omt(P,{Q})/(2\pi i)$ in the space ${\cal H}^\h(\surf)$ is defined by the matrix $- \Bt(\Bt+\q)^{-1}$ (see formula (\ref{opOm})).

The motivation for the definition (\ref{deform-kernels}) is that variational formulas for the bidifferentials $\Omt(P,Q)$ and $\Brt(P,Q)$ defined in this way are similar to variational formulas for the Schiffer and Bergman kernels. Therefore, the deformations $\Omt$ and $\Brt\,,$ analogously to the kernels $\Omega$ and $B\,,$ define rotation coefficients of some Darboux-Egoroff metrics on the Hurwitz space with coordinates $\{\l_k;\lb_k\}$. We find a family of such metrics; they are defined on the Hurwitz space outside the subspace of codimension one given by the equation $\det\,(\Bt+\q)=0\;.$ It turns out that this family also contains a class of metrics which correspond to new Frobenius structures. We call these structures the real doubles of the deformed Frobenius manifolds.

\vspace{0.2cm}
The paper is organized as follows. In the next section we define the Hurwitz spaces and several families of Darboux-Egoroff metrics on them constructed using the bidifferentials introduced above. In Section \ref{SectDeform} we give a definition of Frobenius structures and construct deformations of Dubrovin's Hurwitz Frobenius manifolds \cite{Dubrovin}. In Section \ref{SectDoubles} we construct the real doubles of the deformations. In 
Section \ref{SectG} we compute expressions for the $G$-function on each constructed Frobenius manifold. Section \ref{SectExamples} is devoted to a calculation of prepotentials and $G$-functions of the deformations of Frobenius manifolds and their real doubles in the case of the simplest Hurwitz space in genus one. In that section we also describe the relationship of the example of prepotential, the Chazy equation and isomonodromic deformations related to the Painlev\'e-VI equation. We show that in genus one the constructed one-parameter deformations have a two-parametric ge\-ne\-ra\-li\-za\-tion which can be possibly extended to Hurwitz spaces in any genus which we hope to address in the future.

\section{Darboux-Egoroff metrics on Hurwitz spaces}

\subsection{Hurwitz spaces}
\label{Hurwitz}

Consider a compact Riemann surface $\surf$ of genus $g$ and a meromorphic function $\lambda:\surf\to\C P^1$ of degree $N\,.$ The equation
\beqs
\zeta = \l(P) \;, \qquad P \in \surf 
\eeqs
($\zeta$ is a coordinate on $\C P^1$) represents the surface as an $N$-fold ramified covering of $\C P^1\,.$ The covering is a collection of $N$ copies of $\C P^1$ which are glued together along the cuts connecting the ramification points to form a connected manifold. The ramification points $P_j \in \surf$ are the critical points of the function $\l(P)\;,$ i.e. they satisfy $\l^\prime (P_j) = 0\;;$ their projections $\l_j = \l(P_j) $ on the base of the covering $\C P^1$  are called the branch points.

We assume that the function $\l$ has $m+1$ poles at some points $\infty^0,\dots,\infty^m \in \L\,,$  and we denote by $n_i+1$ the order of the pole at $\infty^i\,.$ In other words, there are $m+1$ points on the covering which project to $\zeta=\infty$ on the base; in the point $\infty^i$  there are $\{ n_i+1 \}$  sheets glued together ($n_0,\dots,n_m \in \N$ are such that  $\sum_{i=0}^m (n_i+1) = N$). The numbers $\{ n_i \}$ are called the ramification indices.
We assume the remaining ramification points which have finite projections on the base, $\l_j < \infty ,$ to be simple ( i.e. there are exactly two sheets glued together at the corresponding point on the covering) and denote their number by $L\,.$

The local parameter near a simple ramification point $P_j \in \L$ (which is not a pole of $\l$) is $x_j(P)= \sqrt{\l(P)-\l_j}$ and in a neighbourhood $P\sim\infty^i$ the local parameter $z_i$ is such that $z_i^{-n_i-1}(P) = \l(P)\,.$

For each genus $g$ of the surface, the Riemann-Hurwitz formula gives the possible values of degree $N$ of the function $\l\,,$ number  $L$ of simple finite branch points and the ramification indices  $n_i$ over infinity:
\beqn
2 g - 2 = - 2N + L + \sum_{i=0}^m n_i \;.
\label{RH}
\eeqn

Two coverings are called equivalent if one of them can be obtained from the other by a permutation of sheets. 
The space of equivalence classes of described coverings is the Hurwitz space; we denote it by $M=M_{g;n_0,\dots,n_m}\;.$ 
We shall work with the following covering $\covM = \covM_{g;n_0,\dots,n_m}$ of the Hurwitz space. A point of the space $\covM$ is a triple $\{ \surf,\l,\{ a_k, b_k \}_{k=1}^g \},$ where $\{ a_k, b_k \}_{k=1}^g$ is a canonical basis of cycles on $\surf\,.$ The branch points $\{\l_i\}$ give a set of local coordinates on the space $\covM\,.$

\subsection{Symmetric bidifferentials on Riemann surfaces}
\label{SectKernels}

On a Riemann surface $\L$ of genus $g$ with a canonical basis of cycles $\{a_k;b_k \}_{k=1}^g\;,\;$ let $\{ \omega_k(P) \}_{k=1}^g$ be the set of holomorphic differentials normalized by $\oint_{a_k}\omega_j=\delta_{jk}\;.$
The symmetric matrix $\B$ of $b$-periods of the surface is defined by $\B_{kj} = \oint_{b_k} \omega_j\;;$ its imaginary part is positive definite.

Now we shall introduce the following bidifferentials on the Riemann surface $\surf\,.$

{\bf 1.} The canonical meromorphic bidifferential $W(P,Q)$ is defined by 
\beqn
\label{W-def}
W(P,Q) := d_\iP d_\iQ \log E(P,Q),
\eeqn
where $E(P,Q)$ is the prime form on the surface. The bidifferential can be uniquely characterized by the following properties: it is symmetric; it has a second-order pole on the diagonal $P=Q$ with biresidue $1\;;$ and its $a$-periods vanish:
\beqn
\oint_{a_k}W(P,Q) = 0 \;, \qquad k = 1,\dots,g \;.
\label{W-aperiods}
\eeqn
The $b$-periods of $W(P,Q)$ are given by the holomorphic normalized differentials:
$\oint_{b_k}W(P,Q) = 2 \pi i \; \omega_k(P) \;,$ $\;k = 1,\dots,g.\;$ 
For a covering $(\L,\l)$ the bidifferential $W$ depends on the simple branch points $\{ \l_j \}$ of the covering according to the Rauch variational formulas \cite{KokKor, Rauch}:
\beqn
\frac{\d W(P,Q)}{\d\l_j}=\frac{1}{2}W(P,P_j)W(Q,P_j)\;,
\label{W-variation}
\eeqn
where $W(P,P_j)$ denotes the evaluation of $W(P,Q)$ at  $Q=P_j$ with respect to the standard local parameter $x_j(Q)=\sqrt{\l(Q)-\l_j}$ near a ramification point $P_j\,:$ 
\begin{equation}
W(P,P_j)=\frac{W(P,Q)}{dx_j(Q)}\Big\vert_{Q=P_j}\;.
\label{notation}
\end{equation}
Being integrated over $b$-cycles of the surface, the Rauch formulas (\ref{W-variation}) give the variational formulas for holomorphic differentials and the matrix $\B$ of $b$-periods:
\beqn
\frac{\d\omega_k(P)}{\d\l_j} = \frac{1}{2} \omega_k(P_j) W(P,P_j) \;, \qquad \frac{\d\B_{kl}}{\d\l_j} = \pi  i \, \omega_k(P_j) \omega_l(P_j) \;.
\label{Rauch}
\eeqn

{\bf 2.} For a covering $(\surf\,, \l)$ of genus $g \geq 1$ consider a symmetric nondegenerate matrix $\q$ which is independent of the branch points $\{ \l_j \}$ and such that the inverse $(\B+\q)^{-1}$ exists. Then, we define a symmetric bidifferential $\Wq(P,Q)$ which is the following deformation of the bidifferential $W(P,Q)\,:$
\beqn
\Wq(P,Q) := W(P,Q)-2 \pi i \sum_{k,l=1}^g(\B+\q)_{kl}^{-1}\omega_k(P)\omega_l(Q)\;.
\label{Wq-def}
\eeqn
This bidifferential has the same singularity structure as the W-bidifferential and satisfies the normalization condition:
\beqn
\oint_{a_k} \Wq(P,Q) + \sum_{j=1}^g (\q^{-1})_{jk} \oint_{b_j} \Wq(P,Q) = 0 \;.
\label{Wq-periods}
\eeqn

The bidifferential $\Wq$ turns into $W\,,$ for example, in the limit when all diagonal entries $\q_{ii}$ of the matrix $\q$ tend to infinity while the off-diagonal entries remain finite. In this limit, the matrix $(\B+\q)^{-1}$ tends to the zero matrix. 

Consider now the Hurwitz space $M_{g;n_0,\dots,n_m}$ of pairs $(\surf\,, \l)\,.$ The equation 
\beqn
\det\,(\,\B+\q\,) = 0
\label{divisor}
\eeqn
defines a divisor in $M_{g;n_0,\dots,n_m}\;,$ which we denote by $\Div\,.$ 
A simple computation shows that $\Wq(P,Q)$ satisfies the variational formulas which formally look exactly as variational formulas (\ref{W-variation}) for $W(P,Q)\,:$
\beqn
\frac{\d \Wq(P,Q)}{\d\l_j}=\frac{1}{2}\Wq(P,P_j)\Wq(Q,P_j) \;.
\label{Wq-variation}
\eeqn
These formulas hold at the points of the Hurwitz space where the bidifferential $\Wq$ is well defined, i.e. outside of divisor $\Div$ (\ref{divisor}).

Note that both bidifferentials $W$ and $\Wq\;,$ as well as the differentials $\omega_k$ and the matrix $\B\;,$ are holomorphic with respect to branch points $\{ \l_k \},$ i.e. they do not depend on $\bar{\l}_k$ (\cite{Fay92}, p. 54). 

{\bf 3.} The Schiffer and Bergman bidifferentials (kernels) are defined on a Riemann surface of genus $g \geq 1$ by
\begin{align}
\Omega(P,Q) &:= W(P,Q) - \pi\sum_{k,l=1}^g(\imb)_{kl}^{-1}\omega_k(P)\omega_l(Q) \;,
\label{Omegadef} \\
B(P,\Qbar) &:= \pi \sum_{k,l=1}^g(\imb)_{kl}^{-1}\omega_k(P)\overline{\omega_l(Q)} \;,
\label{Bdef}
\end{align}
respectively. The following equivalent definitions can be given for these bidifferentials, which, in particular, show that the bidifferentials are independent of the choice of a canonical basis of cycles $\{a_k;b_k\}\,.$ Namely, $\Omega(P,Q)$ can be defined as a symmetric bidifferential having a second order pole with biresidue $1$ at the diagonal $P\sim Q\,,$ such that for any holomorphic differential $\omega$ the following holds: $\iint_\surf \Omega(P,Q)\overline{\omega(Q)} = 0\;.$ The bidifferential $B(P,\Qbar)$ is a regular bidifferential holomorphic  with respect to one of its arguments and antiholomorphic with respect to the other one. The integral operator with the kernel $B(P,\Qbar)/(2\pi i)$ acts in the space $L_2^\h(\surf)$ of $(1,0)$-forms as an orthogonal projector onto the subspace ${\cal H}^\h(\surf)$ of holomorphic $(1,0)$-forms \cite{Fay92}. In particular, in the space ${\cal H}^\h(\surf)$ it acts as the identity operator, i.e. $\iint_\surf B(P,\bar{Q}) \omega(Q)/(2 \pi i) =  \omega(P)\;.$

The periods of bidifferentials (\ref{Omegadef}) and (\ref{Bdef}) are related to each other as follows:
\beqn
\oint_{a_k} \Omega(P,Q) = - \oint_{a_k} B(\bar{P},Q) \;, \qquad \oint_{b_k} \Omega(P,Q) = - \oint_{b_k} B(\bar{P},Q)\;,
\label{SB-periods}
\eeqn
where the integrals are taken with respect to the first argument. The variational formulas for the Schiffer and Bergman kernels have the form:
\begin{equation}
\begin{split}
\frac{\d\Omega(P,Q)}{\d\l_j} = \frac{1}{2} \Omega(P,P_j) \Omega(Q,P_j) \;, 
\qquad
\frac{\d\Omega(P,Q)}{\d\bar{\l}_j} = \frac{1}{2} B(P,\bar{P}_j) B(Q,\bar{P}_j) \;, \\
\frac{\d B(P,\bar{Q})}{\d\l_j} = \frac{1}{2} \Omega(P,P_j) B(P_j,\bar{Q}) \;,
\qquad
\frac{\d B(P,\bar{Q})}{\d\bar{\l}_j} = \frac{1}{2} B(P,\bar{P}_j) \overline{\Omega(Q,P_j)} \;.
\label{SB-variation}
\end{split}
\end{equation}
The notation here is analogous to that in (\ref{notation}), i.e. $\Omega(P,P_j)$ stands for $\left( {\Omega(P,Q)} / {dx_j(Q)} \right) \Big|_{Q=P_j}$ and $B(P,\bar{P}_j) := \left( B(P,\bar{Q}) / \overline{dx_j(Q)} \right) \Big|_{Q=P_j} \;. $

Note that the Schiffer and Bergman kernels depend on both, $\{ \l_k \}$ and $\{ \bar{\l}_k \}$ (holomorphic and anti-holomorphic coordinates on the Hurwitz space), in contrast to bidifferentials $W$ (\ref{W-def}) and $\Wq$ (\ref{Wq-def}), which depend only on holomorphic coordinates $\{\l_k\}\;.$

{\bf 4.} As an analogue of the deformation $\Wq$ (\ref{Wq-def}) of the bidifferential $W\;,$ we shall define deformations of the Schiffer and Bergman kernels, the bidifferentials $\Omt(P,Q)$ and $\Brt(P,Q)\,.$ Let us first introduce the holomorphic differentials
\beqn
v_k(P) := \frac{1}{2 \pi i} \oint_{b_k} \Omega(P,Q) 
\label{v}
\eeqn
and the matrix $\Bt$ of their $b$-periods $\Bt_{kj} = \oint_{b_k}v_j\,:$
\beqs
\Bt :=  \bar{\B}(\bar{\B} - \B )^{-1} \B  \;.
\eeqs
This matrix is symmetric as can be seen from the following representation of $\Bt$ as a sum of two symmetric matrices:
$\Bt = \B(\bar{\B} -\B)^{-1}\B + \B\,.$
Therefore, since $\Bt$ is also anti-Hermitian, it is a purely imaginary matrix. 

The differentials $v_k$ can be characterized as holomorphic differentials on the Riemann surface of genus $g$ whose all $a$- and $b$-periods are purely imaginary except one. Namely, for the differentials (\ref{v}) we have ${\rm Re}\{\oint_{b_j}v_k\}=0$ and  ${\rm Re}\{\oint_{a_j}v_k\}=\delta_{jk}/2\;$ for $\;j,k = 1,\dots, g\,.$ (Recall that by virtue of the Riemann bilinear relations, a holomorphic differential whose all periods are imaginary is zero.) 

\begin{remark} {\rm The differentials given by $a$-periods of the Schiffer kernel, \linebreak $u_k(P):=-\oint_{a_k}\Omega(P,Q)/(2\pi i),$  can be described as holomorphic differentials satisfying the condition ${\rm Re}\{\oint_{b_j}u_k\}=\delta_{jk}/2$ and ${\rm Re}\{\oint_{a_j}u_k\}=0\,.$ }
\end{remark}

The matrix $\Bt$ can also be expressed as the scalar product of the differentials $v_k$  in the space $L_2^\h(\surf)$ of $(1,0)$-forms, i.e. $\Bt_{kl}=\iint_\surf \overline{v_k(P)}\wedge v_l(P)\,.$

The variational formulas for the differentials $v_k$ and the matrix $\Bt$ are analogous to the Rauch formulas (\ref{Rauch}):
\begin{alignat}{3}
\frac{\d v_k(P)}{\d\l_j} &=& \,\frac{1}{2} \, \Omega(P,P_j) v_k(P_j) \;, \qquad 
\frac{\d v_k(P)}{\d\lb_j} &=& \, \frac{1}{2} \, B(P,\bar{P}_j) \overline{v_k(P_j)} \;, 
\label{v-variation}
\\
\frac{\d\Bt_{kl}}{\d\l_j} &=& \pi i \; v_k(P_j) \; v_l(P_j) \;, \;\;\;\qquad
\frac{\d\Bt_{kl}}{\d\lb_j} &=& \pi i \; \overline{v_k(P_j)} \; \overline{v_l(P_j)} \;.
\label{Bt-variation}
\end{alignat}
The following deformed differentials $\Omt$ and $\Brt$ satisfy variational formulas which are similar to those for the kernels $\Omega$ and $B$ (\ref{SB-variation}). Consider a constant nondegenerate matrix $\q$ such that $\q=\q^\iT\;,$  $\bar{\q} = - \q$ and the inverse $(\Bt+\q)^{-1}$ exists. Then, we define
\begin{align}
\Omt(P,Q) &:= \Omega(P,Q) - 2\pi i \sum_{k,l=1}^g(\Bt + \q)_{kl}^{-1} v_k(P) v_l(Q) \;,
\label{Omt-def}
\\
\Brt(P,\Qbar) &:= B(P,\Qbar) - 2\pi i \sum_{k,l=1}^g(\Bt + \q)_{kl}^{-1} v_k(P) \overline{v_l(Q)} \;.
\label{Brt-def}
\end{align}
The bidifferentials $\Omt$ and $\Brt$ turn into the Schiffer and Bergman kernels, respectively, when all entries of the matrix $(\Bt+\q)^{-1}$ tend to zero. This happens, for example, if all diagonal entries of the matrix $\q$ tend to infinity, and all off-diagonal entries remain finite. 

The bidifferentials (\ref{Omt-def}) and (\ref{Brt-def}) are defined for the points of the  Hurwitz space which do not belong to the subspace $\DivR$ of real codimension one given by the equation 
\beqn
\det \,( \, \Bt + \q \, ) = 0 \;.
\label{submanifold}
\eeqn

Similarly to the integral operator with the kernel $B(P,\Qbar)/(2\pi i)\,,$ the integral operator with the deformed kernel $\Brt(P,\Qbar)/(2\pi i)$ also maps the space $L_2^\h(\surf)$ onto ${\cal H}^\h(\surf)\,.$ In the space ${\cal H}^\h(\surf)$ it acts as a linear operator which in the basis $\{v_k\}$ (\ref{v}) is represented by the matrix $\q (\Bt+\q)^{-1}\,.$ Namely, if we denote by ${\bf v}(P)$ the vector of differentials whose $k$-th component is the differential $v_k(P)\;,$ then the following holds:
\beqs
\frac{1}{2\pi i} \iint_\surf \Brt(\bar{P},Q){\bf v}(P) =  \q (\Bt+\q)^{-1} {\bf v}(Q) \;.
\eeqs
The integral operator with the kernel $\Omt(P,Q)/(2\pi i)$ acts in ${\cal H}^\h(\surf)$ as follows:
\beqn
\frac{1}{2\pi i} \iint_\surf \Omt(P,Q)\overline{{\bf v}(P)} = - \Bt(\Bt+\q)^{-1} {\bf v}(Q) \;.
\label{opOm}
\eeqn
Note that when the matrix of parameters $\q$ tends to infinity so that the deformed bidifferential $\Omt$ tends to the Schiffer kernel, the right hand side of (\ref{opOm}) vanishes and this formula turns into the characteristic property of the Schiffer kernel $\Omega\,.$

Periods of the bidifferentials (\ref{Omt-def}) and (\ref{Brt-def}) are related as follows: for any $k=1,\dots,g$
\begin{align}
\begin{split}
\oint_{a_k} \left( \Omt(P,Q) + \Brt(P,\Qbar) \right) &+ \sum_{j=1}^g (\q^{-1})_{kj} \oint_{b_j}  \Omt(P,Q)  =0 \;, \\
\oint_{b_k} \left( \Omt(P,Q) \right. &+ \left. \Brt(P,\Qbar) \right) =0 \;,
\label{periods-rel}
\end{split}
\end{align}
where the integrals are taken with respect to $Q\,.$ 
The following variational formulas for $\Omt$ and $\Brt$ can be derived from (\ref{Omt-def}), (\ref{Brt-def}) by a straightforward computation using variational formulas (\ref{SB-variation}), (\ref{v-variation}) and (\ref{Bt-variation}). They hold outside the subspace $\DivR$ (\ref{submanifold}):
\begin{equation}
\begin{split}
\frac{\d\,\Omt(P,Q)}{\d\l_j} = \frac{1}{2} \Omt(P,P_j) \Omt(Q,P_j) \;, 
\qquad
\frac{\d\,\Omt(P,Q)}{\d\bar{\l}_j} = \frac{1}{2} \Brt(P,\bar{P}_j) \Brt(Q,\bar{P}_j) \;, \\
\frac{\d\Brt(P,\bar{Q})}{\d\l_j} = \frac{1}{2} \Omt(P,P_j) \Brt(P_j,\bar{Q}) \;,
\qquad
\frac{\d\Brt(P,\bar{Q})}{\d\bar{\l}_j} = \frac{1}{2} \Brt(P,\bar{P}_j) \overline{\Omt(Q,P_j)} \;.
\label{q-variation}
\end{split}
\end{equation}
\begin{remark}
{\rm All defined bidifferentials except the Schiffer and Bergman kernels $\Omega$ and $B\,,$ depend on the choice of a canonical basis of cycles $\{a_k;b_k\}\,.$ }
\end{remark}

\subsection{Darboux-Egoroff metrics defined by the bidifferentials}
 \label{SectFlatmetrics}

A diagonal metric ${\bf ds^2}=\sum_i g_{ii}(d\l_i)^2$ is called {\it potential} if there exists a function $U(\{\l_j\})$ whose derivatives give the metric coefficients: $g_{ii} = \d_{\l_i}U$  for any $i\,.$ A metric is called {\it flat} if its curvature tensor vanishes. 
A diagonal potential flat metric is called a {\it Darboux-Egoroff} metric. The Darboux-Egoroff lemma states that a diagonal metric ${\bf ds^2}=\sum_i g_{ii}(d\l_i)^2$ is Darboux-Egoroff if its {\it rotation coefficients} $\beta_{ij}$ defined for $i \neq j$ by
\beqn
\beta_{ij}=\frac{\d_{\l_j}\sqrt{g_{ii}}}{\sqrt{g_{jj}}}
\label{rotation}
\eeqn
are symmetric, $\beta_{ij}=\beta_{ji},$ (this implies ${\bf ds^2}$ is potential) and satisfy the system of equations:
\begin{align}
&\d_{\l_k} \beta_{ij} = \beta_{ik} \beta_{kj} \;,\hsp i,j,k\;\; \mbox{are distinct},
\label{flat1}\\
&\sum_k \d_{\l_k} \beta_{ij} = 0 \qquad \mbox {for all} \;\;\beta_{ij}\;.
\label{flat2}
\end{align}

Consider the Hurwitz space $\covM_{g;n_0,\dots,n_m}$ of coverings $(\L, \l)$ described in Section \ref{Hurwitz}. Let us fix an arbitrary contour $l$ on the surface $\L$ which does not pass through ramification points $\{ P_j \}$ of the covering and whose projection on $\C P^1$ does not change under small variations of the branch points $\{ \l_j \}.$  Let us also fix a function $h(P)$  defined in a neighbourhood of the contour $l\,;$ asume this function to be independent of $\{ \l_j \}.$ Then the following formula defines (see \cite{KokKor}) a family of Darboux-Egoroff metrics on the Hurwitz space:
\beqn
{\bf ds^2} = \sum_{j=1}^L \left( \oint_l h(Q) W(Q,P_j) \right)^2 (d\l_j)^2 \;.
\label{W-metrics}
\eeqn
Following \cite{Dubrovin}, we use the word ``metric" for a bilinear quadratic (not necessary real and positive) form. 

The variational formulas (\ref{W-variation}) for the $W$-bidifferential immediately imply that rotation coefficients for the metrics (\ref{W-metrics}) are given by 
\beqn
\beta_{ij} = \frac{1}{2} W(P_i, P_j) \;,
\label{W-rotation}
\eeqn
where, as usual,  the $W$-bidifferential is evaluated at ramification points with respect to the standard local parameter $x_j(P) = \sqrt{\l(P)-\l_j}\,.$

The following proposition was proven in \cite{KokKor}. Here we reproduce the proof given in \cite{KokKor} since an analogous procedure will be used in our present context. 
\begin{proposition} \cite{KokKor}
\label{Wflat-prop}
Rotation coefficients (\ref{W-rotation}) are symmetric and satisfy equations (\ref{flat1}), (\ref{flat2}) and therefore metrics (\ref{W-metrics}) are the Darboux-Egoroff metrics.
\end{proposition}

{\it Proof.}
The symmetry of the rotation coefficients follows from the symmetry of the bidifferential $W(P,Q)$ with respect to the arguments $P$ and $Q\,.$
Variational formulas (\ref{W-variation}) with $P=P_i,$ $Q=P_k\;,$ for different $i,j,k$ imply relations (\ref{flat1}) for rotation coefficients (\ref{W-rotation}).

To verify relations (\ref{flat2}) let us note that the differential operator $\sum_k \d_{\l_k}$ in (\ref{flat2}) can be represented as follows. Consider a biholomorphic map $(\L,\l) \to (\L,\l+\delta)$ of the covering which takes a point $P$ to the point $P^\delta$ belonging to the same sheet and having projection $\l+\delta$ on the base of the covering. Then, for a function of branch points $f(\{\l_k\})$  we have $\sum_k \d_{\l_k}f = (df^\delta/d\delta)|_{\delta=0}\;,$ where $f^\delta$ is the analog of the function $f$ on the covering $(\L,\l+\delta)\,.$

Note also that the definition of $W(P,Q)$ implies its invariance with respect to the map $(\L,\l) \to (\L,\l+\delta)\,:\;$  if $W^\delta$ is the bidifferential $W$ defined on the covering $(\L,\l+\delta)$
 we have $W(P,Q) = W^\delta(P^\delta,Q^\delta)\,.$ Since the local parameters $x_i(P) = \sqrt{\l(P)-\l_i}$ in neighbourhoods of ramification points also do not change under a simultaneous shift of all branch points and $\l\;,$ we have $\sum_k \d_{\l_k}W(P_i,P_j) = 0\,.$
$\Box$

\vspace{0.3 cm}
Analogously, there exist families of Darboux-Egoroff metrics whose rotation coefficients are given by the other bidifferentials defined above. 
\begin{theorem}
Let a contour $l$ and a function $h$ be as in (\ref{W-metrics}). Then the following formulas define the Darboux-Egoroff  metrics on the Hurwitz space $\covM_{g;n_0,\dots,n_m}$ outside the divisor $\Div$ (\ref{divisor}):
\beqn
{\bf ds^2} = \sum_{j=1}^L \left( \oint_l h(Q) \Wq(Q,P_j) \right)^2 (d\l_j)^2 \;.
\label{Wq-metrics}
\eeqn
The rotation coefficients of metrics (\ref{Wq-metrics}) are given by $\beta_{ij} = \Wq(P_i,P_j)/2\;$ for $\; i \neq j\,.$
\end{theorem}

{\it Proof.}  The proof of Proposition \ref{Wflat-prop} obviously holds for $\beta_{ij} = \Wq(P_i,P_j)/2\,,$ therefore the metrics (\ref{Wq-metrics}) are Darboux-Egoroff. 
$\Box$

The following bilinear quadratic forms were introduced in \cite{doubles}; they can be considered as metrics on the real Hurwitz space, i.e. the moduli space of coverings with local coordinates $\{ \l_k;\lb_k \}.$ We shall denote the real Hurwitz space  by $M_{g;n_0,\dots,n_m}^{\rm real}\;.$ Let now the function $h$ and the projection of the contour $l$ onto the $\l$-sphere be independent of the coordinates $\{ \l_k;\lb_k \}.$ Consider the following two metrics:
\beqn
{\bf ds}^{\bf 2} = \sum_{j=1}^L \left(\oint_lh(Q) \Omega(Q,P_j) \right)^2 (d\l_j)^2 + \sum_{j=1}^L \left( \oint_lh(Q) B(Q,\bar{P_j}) \right)^2 (d\lb_j)^2
\label{SB-metrics1}
\eeqn
and
\beqn
{\bf ds}^{\bf 2} =
{\rm Re} \left\{ 
 \sum_{j=1}^L \left( \oint_lh(Q) \Omega(Q,P_j) + \oint_l \overline{h(Q)}  B(\bar{Q},{P_j}) \right)^2 (d\l_j)^2 \right\} \;.
\label{SB-metrics2}
\eeqn
Both families, (\ref{SB-metrics1}) and (\ref{SB-metrics2}), have rotation coefficients given by
\beqn
\beta_{ij} = \frac{1}{2} \Omega(P_i,P_j) \;, \hsp\hsp
\beta_{i\jb} = \frac{1}{2} B(P_i,\bar{P_j}) \;, \hsp\hsp
\beta_{\ib\jb} = \overline{\beta_{ij}} \;,
\label{SB-rotation}
\eeqn
where $i,j=1,\dots,L$ and the index $\jb$ corresponds to differentiation with respect to $\lb_j\;.$

The proof of the flatness of these metrics is analogous to the proof of Proposition \ref{Wflat-prop}. The variational formulas (\ref{SB-variation}) give relations (\ref{flat1}) for rotation coefficients. To prove relations (\ref{flat2}) we note that all bidifferentials are invariant with respect to the biholomorphic map $(\L,\l) \to (\L,\l+\delta)$ since all of them can be written in terms of $W(P,Q)$ (for example $2\pi i \, \omega_k = \oint_{b_k}W$).
On the space $M^{\rm real}$ (we skip the indices for brevity), equations (\ref{flat2}) read $\sum_{k=1}^\iL(\d_{\l_k} + \d_{\lb_k})\beta_{ij}=0\;;$  to prove them we apply the method of the proof of Proposition \ref{Wflat-prop} with $\delta \in \R$ to the kernels $\Omega$ and $B\,.$ 

Since for finding rotation coefficients and proving the flatness of the metrics (\ref{SB-metrics1})-(\ref{SB-metrics2}) we only used variational formulas for the Schiffer and Bergman kernels, which look identical to those for the bidifferentials $\Omt$ and $\Brt\,,$ the similar metrics can be written in terms of 
$\Omt$ and $\Brt\,.$ Therefore, we have the following theorem. 
\begin{theorem}
Let a contour $l$ and a function $h$ be as in (\ref{SB-metrics1}), (\ref{SB-metrics2}). Then the following formulas define Darboux-Egoroff metrics on the Hurwitz space $\covM_{g;n_0,\dots,n_m}^{\rm real}$ outside the subspace $\DivR$ (\ref{submanifold}):
\beqn
{\bf ds}^{\bf 2} = \sum_{j=1}^L \left(\oint_lh(Q) \Omt(Q,P_j) \right)^2 (d\l_j)^2 + \sum_{j=1}^L \left( \oint_lh(Q) \Brt(Q,\bar{P_j}) \right)^2 (d\lb_j)^2
\label{SBt-metrics1}
\eeqn
and
\beqn
{\bf ds}^{\bf 2} =
{\rm Re} \left\{ 
 \sum_{j=1}^L \left( \oint_lh(Q) \Omt(Q,P_j) + \oint_l \overline{h(Q)}  \Brt(\bar{Q},{P_j}) \right)^2 (d\l_j)^2 \right\} \;.
\label{SBt-metrics2}
\eeqn
The rotation coefficients of metrics of both families (\ref{SBt-metrics1}) and (\ref{SBt-metrics2}) are given by 
\beqn
\beta_{ij} = \frac{1}{2} \Omt(P_i,P_j) \;, \hsp\hsp
\beta_{i\jb} = \frac{1}{2} \Brt(P_i,\bar{P_j}) \;, \hsp\hsp
\beta_{\ib\jb} = \overline{\beta_{ij}} \;. 
\label{SBt-rotation}
\eeqn
\end{theorem}

Note that coefficients of metrics (\ref{SB-metrics1}), (\ref{SB-metrics2}), written in terms of the Schiffer and Bergman kernels, do not depend on the choice of basis of cycles $\{a_k;b_k\}\,.$ Therefore, those metrics are defined on the Hurwitz space $M^{\rm real}\,,$
whereas the metrics (\ref{SBt-metrics1}), (\ref{SBt-metrics2}) are defined on the covering $\covM^{\rm real}$ of the Hurwitz space, i.e. in order to define metrics (\ref{SBt-metrics1}), (\ref{SBt-metrics2}) one must specify the choice of a canonical basis of cycles.

Each family of metrics (\ref{W-metrics}), (\ref{Wq-metrics}), (\ref{SB-metrics1})-(\ref{SB-metrics2}) and (\ref{SBt-metrics1})-(\ref{SBt-metrics2}) contains a class of metrics which correspond to Frobenius structures on the Hurwitz space. Such structures for metrics (\ref{W-metrics}) were found in \cite{Dubrovin} (see also \cite{doubles}). For the family (\ref{SB-metrics1})-(\ref{SB-metrics2}) Frobenius structures were described in \cite{doubles}. In this paper we shall construct Frobenius manifolds corresponding to the metrics (\ref{Wq-metrics}) and (\ref{SBt-metrics1})-(\ref{SBt-metrics2}). Thereby we shall construct deformations of the Hurwitz Frobenius manifolds of \cite{Dubrovin} and \cite{doubles}. 

\subsection{Systems of hydrodynamic type}

A Darboux-Egoroff metric defines (see, for example, \cite{Tsarev}) an integrable system of hydrodynamic type for the branch points $\{ \! \l_k \! \}$ considered as functions of two independent coordinates $x$ and $t:$ 
\beqn
\d_x\l_m = V_m(\{ \l_k \}) \d_t \l_m \;.
\label{hydrodynamic}
\eeqn
where the functions $\{V_m\}\;,$ called the {\it characteristic speeds}, are related to the Christoffel symbols $\Gamma_{nm}^k$ of the metric by:
\beqn
\d_{\l_m} V_n = \Gamma_{nm}^n (V_m-V_n) \;, \qquad m\neq n \;. 
\label{speeds-Christoffel}
\eeqn
The nonvanishing Christoffel symbols for a diagonal metric ${\bf ds^2} = \sum_j g_{jj}(d\l_j)^2$ are given by:
\beqn
\Gamma_{ii}^k = -\frac{1}{2} \frac{\d_{\l_k}g_{ii}}{g_{kk}} \;, \qquad \Gamma_{ii}^i = \frac{1}{2} \frac{\d_{\l_i}g_{ii}}{g_{ii}} \;, \qquad \Gamma_{ij}^i = \frac{1}{2} \frac{\d_{\l_j}g_{ii}}{g_{ii}} \qquad i,j,k \;\; \mbox{are distinct} \;.
\label{diagChristoffel}
\eeqn
If the metric ${\bf ds^2}$ is Darboux-Egoroff, then the equations (\ref{speeds-Christoffel}) for characteristic speeds are compatible. In particular, for the metrics (\ref{W-metrics}) the systems of hydrodynamic type (\ref{hydrodynamic}) were constructed and solved in \cite{KokKor}. Here we note that analogous systems are associated with the Darboux-Egoroff metrics (\ref{Wq-metrics}). Namely, if the metric ${\bf ds^2}$ belongs to the family (\ref{Wq-metrics}), corresponding to the bidifferential $\Wq\;,$ then the system (\ref{speeds-Christoffel}) is defined on the Hurwitz space outside the divisor $\Div$ (\ref{divisor}). Solutions to (\ref{speeds-Christoffel}) are given by
\beqn
V_m (\{\l_k\}) = \frac{\oint_{l_1} h_1(Q) \Wq(Q,P_m)}{\oint_l h(Q) \Wq(Q,P_m)} \;; 
\label{Vm}
\eeqn
where the contour $l$ and function $h$ are those which define the metric ${\bf ds^2}$ as in (\ref{W-metrics}); and $l_1$ and $h_1$ are such that the projection of the contour $l_1$ on the base of the covering  and the function $h_1$ are independent of branch points $\{ \l_j \}.$ Relations (\ref{speeds-Christoffel}) for the functions (\ref{Vm}) can be verified by a simple calculation using the variational formulas (\ref{Wq-variation}) for the bidifferential $\Wq\,.$

Solutions to the system of hydrodynamic type (\ref{hydrodynamic}) are constructed by the generalized {\it hodograph method} \cite{Tsarev}. Namely, 
for the functions $V_m(\{\l_m\})$ which satisfy equations (\ref{speeds-Christoffel}) consider an arbitrary solution $\{U_m(\{\l_k\})\}$ to the system
\beqn
\frac{\d_{\l_n}U_m}{U_m-U_n} = \frac{\d_{\l_n}V_m}{V_m-V_n}\;, \qquad m,n = 1,\dots L \;. 
\label{W-hodograph}
\eeqn
Then, the system of equations
\beqn
U_m(\{\l_k\}) = t + V_m (\{\l_k\})  \, x
\label{hodograph}
\eeqn
defines an implicit solution $\{\l_m(x,t)\}$ to the system of hydrodynamic type (\ref{hydrodynamic}). 
A solution to the system (\ref{W-hodograph}) is obviously given by formulas (\ref{Vm}) with some other pair $(l_2\,,h_2)$ instead of $(l_1\,,h_1)\,.$
\pagebreak[1]

Let us assume $l_1=l_2=l\,.$ Then, the hodograph method for the system  (\ref{hydrodynamic}), (\ref{Vm}) is summarized in the following theorem. 
\begin{theorem}
Let us fix a contour $l$ on the covering which does not pass through ramification points. Consider functions $h,\;h_1,\;h_2$ defined in a neighbourhood of the contour.  Assume that the functions and the projection of the contour $l$ on $\C P^1$  are independent of the branch points $\{\l_j\}\;.$
Then, a solution $\{\l_m(x,t)\}$ to the system of hydrodynamic type (\ref{hydrodynamic}), (\ref{Vm}) can be implicitly defined on $M_{g;n_0,\dots,n_m} \setminus \Div$ (where $\Div$ is the divisor (\ref{divisor})) by the following system:
\beqs
\oint_{l} \left( h_2(Q) - h(Q)t - h_1(Q)x \right) \Wq(Q,P_m) = 0\;, \qquad m=1,\dots, L \;.
\eeqs
\end{theorem}

For families (\ref{SB-metrics1})-(\ref{SB-metrics2}), (\ref{SBt-metrics1})-(\ref{SBt-metrics2}) of Darboux-Egoroff metrics a naive definition, analogous to (\ref{hydrodynamic}), of systems of hydrodynamic type does not lead to a compatible system on variables $\{\l_k\}$ and $\{ \lb_k \}.$ In this case the equations on $\l_k$ and $\lb_k$ are not complex conjugate to each other. However, an analogous procedure may work in the sense of analytic continuation, if $\l_k$ and $\lb_k$ are considered as independent complex variables.

\section{Deformations of Hurwitz Frobenius structures}
\label{SectDeform}

\subsection{Definition of Frobenius manifold}

\begin{definition}
A commutative associative algebra over $\C$ with a unity ${\bf e}$ is called a {\bf Frobenius algebra} if
it is supplied with a $\C$-bilinear symmetric nondegenerate inner product $\langle\cdot,\cdot\rangle$ which has the property $\langle {\bf x}\cdot {\bf y},{\bf z}\rangle =\langle {\bf x},{\bf y}\cdot {\bf z}\rangle$ for arbitrary elements ${\bf x},{\bf y},{\bf z}$ of the algebra.  
\end{definition}
\begin{definition}
$M$ is a {\bf Frobenius manifold} of charge $\nu$ if a structure of a Frobenius algebra is defined in any tangent plane $T_{\chi}M\,;$ this structure should smoothly depend on the point $\chi \in M$ and be such that
\enumerate
\item[{\bf F1}] the inner product $\langle\cdot ,\cdot\rangle $ is a flat metric on $M$ (not necessarily real positive definite);
\item[{\bf F2}] the unit vector field ${\bf e}$ is covariantly constant with respect to the Levi-Civita connection $\nabla$ of the metric $\langle\cdot ,\cdot\rangle\;,$ i.e. the covariant derivative in the direction of any vector field ${\bf x}$ on $M$ vanishes: $\nabla\!_{\bf x} \,{\bf e}=0\,;$
\item[{\bf F3}] the tensor $(\nabla_{\bf w}{\bf c})({\bf x},{\bf y},{\bf z})$ is symmetric in four vector fields ${\bf x},{\bf y},{\bf z},{\bf w}$ on $M\;,$ where ${\bf c}$ is the following symmetric $3$-tensor: ${\bf c}({\bf x},{\bf y},{\bf z})=\langle {\bf x}\cdot {\bf y},{\bf z}\rangle\;;$
\item[{\bf F4}] there exists a vector field $E$ (the Euler vector field)   such that for any pair of vector fields ${\bf x}$ and ${\bf y}$ on M
\beqn
\nabla\!_{\bf x}(\nabla\!_{\bf y}\, E)=0\;,
\label{covlin}
\eeqn
\beqn
[E,{\bf x}\cdot {\bf y}]-[E,{\bf x}]\cdot {\bf y}- {\bf x}\cdot[E,{\bf y}]={\bf x}\cdot {\bf y}\;,
\label{Euler1}
\eeqn
\beqn
\Lie_\iE \langle {\bf x},{\bf y} \rangle := E \langle {\bf x},{\bf y} \rangle - \langle [E,{\bf x}],{\bf y} \rangle - \langle {\bf x},[E,{\bf y}] \rangle = (2-\nu) \langle {\bf x},{\bf y} \rangle \;.
\label{Euler2}
\eeqn
\label{defFrob}
\end{definition}

The structure described in Definition \ref{defFrob} is equivalent to the WDVV system (\ref{WDVV})-(\ref{quasihomogeneity}). Requirement {\bf F3} implies the existence of a function $F$ depending on flat coordinates $t= \{ t^\iA \}$ of the metric from {\bf F1} whose third order derivatives give the tensor ${\bf c}\,:$
\beqn
\frac{\d^3 F(t)}{\d_{t^A}\d_{t^B}\d_{t^C}}={\bf c}(\d_{t^A},\d_{t^B},\d_{t^C})= \langle \d_{t^A}\cdot\d_{t^B},\d_{t^C} \rangle\;.
\label{prepot-def}
\eeqn
The associativity conditions of the Frobenius algebra are equivalent to the equations (\ref{WDVV}) and the existence of the vector field $E$ from {\bf F4} provides the quasihomogeneity (\ref{quasihomogeneity}) for the function $F\,.$

The function F defined by (\ref{prepot-def}) up to a quadratic polynomial in flat coordinates is called the {\it prepotential} of the Frobenius manifold $M\,.$
\begin{definition}
A Frobenius manifold $M$ is called {\bf semisimple} if for any point $\chi \in M$ the Frobenius algebra in the tangent space $T_{\chi}M$ has no nilpotents.
\end{definition}

For semisimple Frobenius manifolds, the flat metric in the definition of a Frobenius manifold is also diagonal and potential (\cite{Dubrovin}, Lemmas 3.6-3.7), hence it is in fact a Darboux-Egoroff metric.

In this paper we only consider semisimple Frobenius manifolds.

\subsection{Flat metrics}
\label{FlatFrob}

The Frobenius structures on Hurwitz spaces which correspond to the Darboux-Egoroff metrics of the type (\ref{W-metrics})  were found by Dubrovin \cite{Dubrovin}. In \cite{doubles} the construction of \cite{Dubrovin} was reformulated in terms of the bidifferential $W(P,Q)$ (\ref{W-def}). Analyzing this construction one can see that it is essentially based on the following properties of $W(P,Q)\;.$
\begin{itemize}
\item The variational formulas (\ref{W-variation}) for $W(P,Q)$ which provide the flatness for the metrics (\ref{W-metrics}).
\item Invariance of $W(P,Q)$ with respect to two maps of coverings: $(\L,\l) \to (\L,\l+\delta) $ and 
 $(\L,\l) \to (\L,(1+\epsilon)\l)$ which take a point $P$ of the surface to the points $P^\delta$ and $P^\epsilon$ which lie on the same sheet of the covering and have projections $\l+\delta$ and $(1+\epsilon)\l$ on $\C P^1\,,$ respectively. The bidifferential $W(P,Q)$ is invariant under the action of these two maps, i.e. we have
$ W^\delta(P^\delta,Q^\delta) = W(P,Q)$ and $ W^\epsilon(P^\epsilon,Q^\epsilon) = W(P,Q) \;, $ where $W^\delta$ and $W^\epsilon$ are the bidifferentials $W$ defined on the corresponding coverings.

These properties provide the validity of conditions ({\bf F2}) and ({\bf F4}) for a certain class of the metrics (\ref{W-metrics}).
\item The type of singularity of $W(P,Q)$ at $P\simeq Q$ (quadratic pole with biresidue $1$).
\item The normalization $\oint_{a_k}W(P,Q)=0$ for all $k=1,\dots,g\,.$
\end{itemize}

\vspace{0.4cm}
Let us notice that the bidifferential $\Wq(P,Q)$ (\ref{Wq-def}) possesses a similar set of  properties. The variational formulas (\ref{Wq-variation}) for $\Wq(P,Q)$ are identical to those for $W(P,Q)\,.$ Furthermore, $\Wq(P,Q)$ is invariant with respect to the maps $(\L,\l) \to (\L,\l+\delta) $ and  $(\L,\l) \to (\L,(1+\epsilon)\l)$ since it is expressed in terms of the bidifferential $W(P,Q)\,,$ holomorphic normalized $1$-forms $\{ \omega_k \}_{k=1}^g$ and the matrix of $b$-periods $\B\,.$ Finally, $\Wq(P,Q)$ has the same singularity structure as $W(P,Q)$ at $P\simeq Q$ and is normalized by (\ref{Wq-periods}).

Therefore, we conclude that in analogy with the construction of \cite{Dubrovin} it should be possible to find Frobenius structures for Darboux-Egoroff metrics from the family (\ref{Wq-metrics}). Then $g(g+1)/2$ parameters contained in the bidifferential $\Wq(P,Q)$ will be inherited by the corresponding Frobenius manifolds.

Consider now the limit in which some of the entries of the matrix $\q$ tend to infinity in such a way that for {\it any} matrix $\B$ independent of $\q$ the matrix $(\B+\q)^{-1}$ tends to the zero matrix (for example, let $\q_{ii} \to \infty$ for any $i$ and $\q_{ij}$ be finite for $i\neq j$). In this limit the bidifferential $\Wq$ turns into $W\,,$ and our construction coincides with that of \cite{Dubrovin}. For a finite constant symmetric matrix $\q$ it gives a $g(g+1)/2$-parametric deformation of Frobenius manifolds of \cite{Dubrovin}.

Each matrix $\q$ defines by the equation (\ref{divisor}) the divisor $\Div$ on the Hurwitz space $\covM=\covM_{g;n_0,\dots,n_m}\;.$ We shall describe  structures of the Frobenius manifolds corresponding to some metrics of the type (\ref{Wq-metrics}). These Frobenius structures are defined on the Hurwitz space outside the divisor $\Div\,.$ 

The associative algebra is defined on each tangent space by
\beqn
\d_{\l_i} \cdot \d_{\l_j} := \delta_{ij} \d_{\l_i} \;;
\label{multiplication}
\eeqn
the coordinates $\{\l_j\}$ are thus canonical for multiplication. As is easy to see, the algebra (\ref{multiplication}) does not have nilpotents. The unit vector field is given by
\beqn 
{\bf e} = \sum_{i=1}^L \d_{\l_i} \;.
\label{e}
\eeqn
For this multiplication a bilinear quadratic form $\langle, \rangle$  has the property $\langle x\cdot y, z\rangle = \langle x,y\cdot z\rangle$ if it is diagonal in the coordinates $\{\l_j\}\,.$ Therefore the metrics (\ref{Wq-metrics}) define a Frobenius algebra in the tangent space at each point of the Hurwitz space outside the divisor $\Div\,.$

The Euler vector field has the following standard form:
\beqn
E := \sum_{i=1}^L \l_i \d_{\l_i} \;.
\label{Euler}
\eeqn
It is easy to see that condition (\ref{Euler1}) is satisfied for the multiplication (\ref{multiplication}). The condition (\ref{Euler2}) for a diagonal metric ${\bf ds^2}=\sum_i g_{ii} (d\l_i)^2$ reduces to $E(g_{jj})=-\nu g_{jj} \;.$ 
To verify the requirement $\nabla\!_{\bf x} {\bf e} =0$ (${\bf F2}$) we note that the metrics (\ref{Wq-metrics}) are potential, i.e. $\d_{\l_{j}}g_{ii} = \d_{\l_i}g_{jj},$ and therefore, as is easy to check by a straightforward calculation, $\nabla\!_{\bf x} {\bf e} =0$ holds if ${\bf e}(g_{jj})=0\;.$ Thus, among the metrics (\ref{Wq-metrics}), we need to find those which for some constant $\nu$ satisfy 
\beqn
{E}(g_{jj}) = - \nu g_{jj}  \qquad
{\mbox {and}} \qquad {\bf e}(g_{jj}) =0\;.
\label{relations}
\eeqn
The action of the vector fields ${\bf e}$ and $E$ on a function of the canonical coordinates $\{ \l_j \}$ only can be represented via the maps of coverings: $(\L,\l) \to (\L,\l+\delta)$ and $(\L,\l) \to (\L,(1+\epsilon)\l)\,,$ respectively. These maps take a point $P$ of the surface to the points $P^\delta$ and $P^\epsilon$ which lie on the same sheet of the covering and have projections $\l+\delta$ and $(1+\epsilon)\l$ on $\C P^1$ (i.e. $\l(Q^\delta) = \l(Q)+\delta$ and $\l(Q^\epsilon) = (1+\epsilon)\l(Q)$). The bidifferential $\Wq(P,Q)$ is invariant under the action of these two maps, i.e. we have $\Wq^\delta(P^\delta,Q^\delta) = \Wq(P,Q) $ and $\Wq^\epsilon(P^\epsilon,Q^\epsilon) = \Wq(P,Q) \;,
$ 
where $\Wq^\delta$ and $\Wq^\epsilon$ are the bidifferentials $\Wq$ defined on the corresponding coverings.
For the evaluation of $\Wq(P,Q)$ at $P=P_j$ we have to take into account transformations of the standard local parameter  near a ramification point: $x_j^\delta(P^\delta) = x_j(P)$ and $x_j^\epsilon(P^\epsilon) = \sqrt{1+\epsilon} \; x_j(P)\,.$

Then it is easy to see that the requirement $E(g_{jj}) = - \nu g_{jj}$ is satisfied for a metric of the type (\ref{Wq-metrics}) if $h(Q) = const\;\l^n(Q)$ and the contour $l$ is invariant under the map $\l\to(1+\epsilon)\l\,,$ i.e. if it is either a closed contour or a contour connecting points $\infty^i$ and $\infty^j\,:$ 
\begin{multline}
E(g_{jj}) = \frac{d}{d\epsilon}\Big|_{\epsilon=0}
\left( \oint_{l^\epsilon} \l^n(Q^\epsilon) \frac{\Wq^\epsilon(Q^\epsilon,P^\epsilon)}{dx^\epsilon_j(P^\epsilon)} \Big|_{P=P_j} \right)^2 \\
= \frac{d}{d\epsilon}\Big|_{\epsilon=0}
(1 + \epsilon)^{2n-1} \left( \oint_{l} \l^n(Q) \frac{\Wq(Q,P)}{\;dx_j(P)} \Big|_{P=P_j} \right)^2 = (2n-1)g_{jj}\;.
\label{temp1}
\end{multline}
The condition ${\bf e}(g_{jj})=0$ holds if the combination of a contour $l$ and a function $h(Q)=const\;\l(Q)$ in (\ref{Wq-metrics}) is one of the following combinations. Let us write these combinations in the form of integral ope\-ra\-tions applied to some $(1,0)$-form $f(Q)$ on the surface:
\begin{alignat*}{2}
{\bf 1.\;\;} & \I_{t^{i;\a}} [f(Q)] := \frac{1}{\a} \; \underset{\infty^i} {\res} \;  \l(Q)^\frac{\a}{n_i+1} f(Q)
 \qquad &  i&=0, \dots, m\;; \; \a=1, \dots, n_i \;.\\
{\bf 2.\;\;} & \I_{v^i} [f(Q)] := 
\; \underset{\infty^i} {\res} \; \l(Q)f(Q) & \qquad   i&=1,\dots,m \;.\\
{\bf 3.\;\;} & \I_{w^i} [f(Q)] := \mathrm{v.p.} \int_{\infty^0}^{\infty^i} f(Q) & \qquad i&=1,\dots,m\;. \\
{\bf 4.\;\;} & \I_{r^k} [f(Q)] := - \oint_{a_k} \l(Q) f(Q) - \sum_{n=1}^g(\q^{-1})_{nk} \oint_{b_n} \l(Q) f(Q) & \qquad k&=1, \dots, g\;.\\
{\bf 5.\;\;} & \I_{s^k} [f(Q)] := \frac{1}{2\pi i} \oint_{b_k} f(Q) & \qquad k&=1,\dots,g\;.
\end{alignat*}
Here the principal value near infinity is defined by omitting the divergent part of an integral as a function of the local parameter $z_i$  (such that $ \l=z_i^{-n_i-1})\,.$ The number of operations is $L=  \sum_{i=0}^m n_i  +  2m + 2g\,,$ where $\sum_{i=0}^m (n_i  +  1) = N\,,$ according to the Riemann-Hurwitz formula (\ref{RH}).

We shall denote the set of operations 1.-5. by $\{ \I_{t^\iA} \},$ i.e. we define $t^\iA \in \{ t^{i;\a} \;; \; v^i \;, w^i \;; \; r^k \;, s^k \}.$
Here, $t^\iA$ is used as a formal index, however, later it will denote a flat coordinate of the flat metric of a Frobenius manifold. 
\begin{theorem}
\label{thm-primary}
Let us choose a point $P_0$ on the surface which is mapped to zero by the function $\l\,,$ i.e.  $\l(P_0)=0\;,$ and let all basic contours $\{a_k,b_k\}$ on the surface start at this point. Let the constant matrix $\q$ be symmetric nondegenerate and such that $\det(\B+\q)\neq 0\;.$ Then, the operations $\I_{t^\iA}$ applied to $f_\iP(Q):=\Wq(P,Q)$ give a set of $L$ differentials, 
called primary, whose characteristic properties are listed below. 
\begin{alignat*}{4}
& \;\; {\bf {\mbox \bf {Primary}}} && {\bf {\mbox \bf {differential}}}  && \hsp {\bf {\mbox \bf {Characteristic}}} \;\;\; {\bf{\mbox \bf {property}}} && \\
&{\bf 1.}\;\; \phi_{t^{i;\a}}(P) &:=&\; \I_{t^{i;\a}}[\Wq(P,Q)] && \hsp \sim z_i^{-\a-1}(P)dz_i(P) \;,  \; P \sim \infty^i\;;  \;\; &i& = 0,  ... ,m \;; \\
&&&&& \hsp {\mbox{single valued on }} \surf \;;  &\a&=1, ... , n_i\;. \\
&{\bf 2.}\;\;  \phi_{v^i}(P) &:=&\; \I_{v^i}[\Wq(P,Q)] && \hsp \sim -d\l(P) \;, \;\; P \sim \infty^i \;; & \hsp  i&=1,\dots,m\;. \\
&&&&& \hsp {\mbox{single valued on }} \surf \;; \\
&{\bf 3.}\;\;  \phi_{w^i}(P) &:= &\; \I_{w^i}[\Wq(P,Q)]: && \hsp \underset{\infty^i}{\res} \; \phi_{w^i} = 1 \;; \;\; \underset{\infty^0}{\res} \; \phi_{w^i} = -1 \;; & \hsp i&=1,\dots,m\;. \\
&&&&& \hsp {\mbox{single valued on }} \surf \;; \\
&{\bf 4.}\;\;   \phi_{r^k}(P) &:= &\; \I_{r^k}[\Wq(P,Q)]: &&\hsp{\mbox{has no poles,}}& \hsp k&=1, \dots, g \;.\\
&&&&& \hsp \phi_{r^k}(P^{a_j}) - \phi_{r^k}(P) = -2 \pi i (\q^{-1})_{kj} d\l(P) \;, & \\
&&&&&  \hsp \phi_{r^k}(P^{b_j}) - \phi_{r^k}(P) = \delta_{kj}2 \pi i d\l(P) \;; \\
&{\bf 5.}\;\;   \phi_{s^k}(P) &:= &\; \I_{s^k}[\Wq(P,Q)]: && \hsp \mbox{holomorphic differential on $\surf\,.$ } & \hsp k&=1,\dots,g \;.
\end{alignat*}
Here $z_i$ is the local parameter near $\infty^i$ such that $z_i^{-n_i-1}=\l\;;$ $n_i$ is the ramification index at $\infty^{i}\,;$  
$\phi (P^{a_j})-\phi(P)$ and $\phi (P^{b_j})-\phi(P)$ denote the transformations of a differential $\phi$ under analytic continuation along cycles $a_j$ and $b_j\,,$ respectively. 

The primary differentials {\bf 1.}-{\bf 5.} satisfy the following normalization condition ($\delta$ is the Kronecker symbol):
\beqn
\oint_{a_k}\phi_{t^\iA} + \sum_{n=1}^g(\q^{-1})_{nk} \oint_{b_n} \phi_{t^\iA}=\delta_{t^A,s^k}\;.
\label{primary-periods}
\eeqn

Let $\phi$ be one of the primary differentials. Then, the following metrics 
\beqn{\bf ds}_\phi^{\bf 2} = \frac{1}{2} \sum_{i=1}^L \phi^2(P_i) (d\l_i)^2 
\label{phi-metrics}
\eeqn
belong to the family (\ref{Wq-metrics}). Their diagonal entries $g_{ii} = \phi^2(P_i)/2$  satisfy the relation ${\bf e}(g_{ii})=0\,.$
\end{theorem}

We shall denote the set of differentials by $\{ \phi_{t^\iA} \},$ i.e. we assume that the index $t^\iA$ belongs to the set of indices $\{ t^{i;\a} \;; \; v^i \;, w^i \;; \; r^k \;, s^k \}\,.$

{\it Proof.} According to the assumption made in the theorem, the cycles $a_k$ and $b_k$ intersect each other at the point $P_0$ such that $\l(P_0)=0\,.$ Therefore, as can be verified by a simple local calculation in a neighbourhood of the point $P_0\,,$ the order of integration can be changed in the integral $\oint_{a_k}\oint_{b_k}\l(P)\Omega(P,Q)\;.$ Similarly, one can prove that the following change of order of integration is valid:
\beqn
\oint_{a_k}\!\! \I_{t^\iA}\![\Wq(P,Q)] = \I_{t^\iA} \!\! \left[ \oint_{a_k} \!\! \Wq(P,Q) \right] + \delta_{t^\iA,s^k} \;; 
\;\;\;\;
\oint_{b_k}\!\! \I_{t^\iA}[\Wq(P,Q)] = \I_{t^\iA} \!\! \left[ \oint_{b_k}\!\!\Wq(P,Q) \right] .
\label{oper-commut}
\eeqn
Therefore, the normalization (\ref{Wq-periods}) of $\Wq(P,Q)$ implies the normalization (\ref{primary-periods}) of the primary differentials.

Now we shall use the invariance of the bidifferential $W$ under the biholomorphic map of co\-ve\-rings $(\surf\,, \l) \to (\surf\,, \l+\delta)$ to prove that the unit vector field annihilates coefficients of the metrics (\ref{phi-metrics}). 
The action of the vector field ${\bf e}$ on $\phi(P_j)$ is given by  the derivative $(d/d\delta)\phi^\delta(P^\delta_j)|_{\delta=0},$ where $\phi^\delta$ and $P^\delta$ are the corresponding objects on the covering $(\L,\l+\delta)\,.$
For the primary differential  $\phi_{t^{i;\a}}$ we have: 
\begin{align*} 
{\bf e}(\phi_{t^{i;\a}}(P_j)) &= \frac{d}{d\delta}\Big|_{\delta=0}
\left\{ \phi^\delta_{t^{i;\a}}(P_j^\delta) \right\} = \frac{d}{d\delta}\Big|_{\delta=0} \left\{ \frac{1}{\a} \; \underset{\infty^i}{\res} (\l(P)+\delta)^{\frac{\a}{n_i+1}} \Wq^\delta(P^\delta,P^\delta_j) \right\} \\
& = \frac{d}{d\delta}\Big|_{\delta=0} \left\{ \frac{1}{\a} \; \underset{\infty^i}{\res} \left( z_i^{-\a}(P) + \frac{\a}{n_i+1} (z_i(P))^{-\a + n_i + 1} \delta + {\cal O}(\delta^2) \right)  \Wq(P,P_j) \right\} 
\\
&= \frac{1}{n_i+1} (z_i(P))^{-\a + n_i + 1} \Wq(P,P_j) \;,
\end{align*}
which is zero for $\a = 1,\dots,n_i+1 $ (for $\a=n_i+1$ this computation shows that ${\bf e}(\phi_{v^i}(P_j))=0$). For primary differentials $\phi_{w^i}$ and $\phi_{s^k}$ the relation ${\bf e}(\phi(P_j))=0$ follows from the invariance of $\Wq(P,Q)$ and the path of integration under the map $\l \to \l+\delta\,.$ For $\phi=\phi_{r^k}$ this relation easily follows from the vanishing of the combination of periods (\ref{Wq-periods}) for $\Wq(P,Q)\,.$
$\Box$

Thus, we have $L$ (see the Riemann-Hurwitz formula (\ref{RH})) Darboux-Egoroff metrics (\ref{phi-metrics}) which satisfy the requirements of the definition of a Frobenius manifold. 

The next lemma shows that one uniquely specifies a holomorphic differential by fixing the values of combinations of its periods which appear in the right hand side in (\ref{primary-periods}) .
\begin{lemma} 
\label{lemma-norm}
Let $\surf$ be a Riemann surface and $\B$ be its matrix of $b$-periods. Consider a constant symmetric nondegenerate matrix $\q$ such that the sum $(\B + \q)$ is also nondegenerate. 
Then a holomorphic differential $\,\omega$ on the surface $\surf$ vanishes if for every $k=1,\dots,g$
\beqn
\oint_{a_k}\omega + \sum_{n=1}^g(\q^{-1})_{nk} \oint_{b_n} \omega = 0 \;.
\label{combination}
\eeqn
\end{lemma}

{\it Proof.} A holomorphic differential $\omega$ can be represented as a linear combination of the holomorphic differentials $\omega_k$ normalized by the condition $\oint_{a_j}\omega_k = \delta_{jk}\;.$ Then, the lemma can be proved by a simple calculation using the well-known fact that a holomorphic differential vanishes if all its $a$-periods vanish. 
$\Box$
\begin{proposition}
\label{Prop-unique}
 Let $w$ be a differential on the Riemann surface having only poles with a given singular part and (or) a given non-singlevaluedness of additive type along basic cycles. Then the differential $w$ can be uniquely fixed by specifying the values of the combinations $\oint_{a_k}w + \sum_{n=1}^g(\q^{-1})_{nk} \oint_{b_n}w$ of its periods for each $k=1,\dots,g\,,$ where the constant symmetric matrix $\q$ is such that $\det(\B+\q)\neq 0 \;.$
\end{proposition}

{\it Proof.} Suppose there exist two differentials with identical singularity structures of the type described in the proposition. Then, their difference is zero by virtue of Lemma \ref{lemma-norm}.
$\Box$

\subsection{Flat coordinates}
\label{FlatCoord}

For a flat metric there exists a set of {\it flat coordinates}.
These are coordinates in which coefficients of the metric are constant.  
The Christoffel symbols in flat coordinates vanish and the covariant derivative $\nabla_{t^\iA}$ along the vector  field in the direction of the flat coordinate $t^\iA$  coincides with the usual partial derivative $ \d_{t^\iA}\,.$ Therefore, flat coordinates can be found from equations $\nabla\!_{\bf x}\nabla\!_{\bf y} \, t = 0$ where  ${\bf x}$ and ${\bf y}$ are arbitrary vector fields on the manifold. The next theorem shows that flat coordinates of the metric (\ref{phi-metrics}) can be found by applying the operations $\I_{t^\iA}$ to the primary differential $\phi$ which defines the metric. 
\begin{theorem}
\label{thm_flatcoord}
The following functions form a set of flat coordinates of the metric ${\bf ds}_\phi^{\bf 2}$ (\ref{phi-metrics}):
\begin{alignat*}{5}
& t^{i;\a} &:=& -(n_i+1) &\;& \I_{t^{i;1+n_i-\a}} [\phi] = \frac{n_i+1}{\alpha-n_i-1} \; \underset{\infty^i}{\res} \; z_i^{\alpha-n_i-1} \phi \qquad     
&& i = 0, \dots, m \;; \; \a = 1, \dots, n_i \\
& v^i &:=& - \I_{w^i} [\phi] &=& - {\rm v.p.}\int_{\infty^0}^{\infty^i} \phi
&& i  = 1, \dots, m \\
& w^i &:=& - \I_{v^i} [\phi] &=& - \; \underset{\infty^i} {\res} \; \l\phi
&& i  = 1, \dots, m \\
& r^k &:=&\; \I_{s^k} [\phi] &=& \; \frac{1}{2\pi i} \oint_{b_k} \phi
&& k  = 1, \dots, g \\
& s^k &:=&\; \I_{r^k} [\phi] &=& - \oint_{a_k} \l\phi - \sum_{n=1}^g (\q^{-1})_{kn} \oint_{b_n} \l\phi 
&& k  = 1, \dots, g \;.
\end{alignat*}
As before, we denote the above functions by $\{t^\iA\}\,,$ i.e. we assume $t^\iA \in \{ t^{i;\a} \;; \; v^i \;, w^i \;; \; r^k \;, s^k \}\,.$
\end{theorem}

{\it Proof.}
Let us verify that the functions $\{t^\iA\}$ satisfy equations $\nabla\!_{\bf x}\nabla\!_{\bf y} \, t = 0$ defining flat coordinates of the metric ${\bf ds}_\phi^{\bf 2}\,.$ These equations can be rewritten for the basis vector fields ${\bf x},{\bf y} \in \{\d_{\l_j}\}$  in canonical coordinates $\{\l_k\}$ as follows: 
\beqn
\d_{\l_i}\d_{\l_j} t = \sum_{k=1}^L \Gamma^k_{ij} \d_{\l_k} t \;, \qquad i,j = 1,\dots,L \;,
\label{flat-sys}
\eeqn
where $\Gamma^k_{ij}$ denote the Christoffel symbols for the Levi-Civita connection of the metric ${\bf ds}_\phi^{\bf 2}\,.$ 

The variational formulas (\ref{Wq-variation}) for $\Wq(P,Q)$ imply the following expressions for derivatives of primary differentials:
\beqn
\frac{\d\phi_{t^A}(P)}{\d\l_j} = \frac{1}{2} \phi_{t^A}(P_j) \Wq(P,P_j) \;.
\label{primary-variation}
\eeqn
Using (\ref{primary-variation}) we find the nonzero Christoffel symbols for the diagonal metric ${\bf ds}_\phi^{\bf 2}$ in terms of the primary differential $\phi\,:$
\beqs
\Gamma_{ik}^k =  \beta_{ik} \frac{\phi(P_i)}{\phi(P_k)} = - \Gamma_{ii}^k  \qquad {\mbox{for }} k\neq i \;; \;\;\mbox{ and}  
\qquad \Gamma_{kk}^k = - \sum_{j,j\neq k} \Gamma_{kj}^k \;.
\eeqs
To prove the last equality one uses the fact that the unit vector field annihilates coefficients of the metric ${\bf ds}_\phi^{\bf 2}$ (\ref{phi-metrics}).
Then, system (\ref{flat-sys}) takes the form:
\begin{align}
\d_{\l_i}\d_{\l_j} t &= \beta_{ij} \left( \frac{\phi(P_j)}{\phi(P_i)} \d_i t + \frac{\phi(P_i)}{\phi(P_j)} \d_j t  \right) \;, \qquad i\neq j \;,
\label{flatcoord1}\\
{\bf e}(t) &= {\mbox const} \;.
\label{flatcoord2}
\end{align}
To show that the system (\ref{flatcoord1})-(\ref{flatcoord2}) is equivalent to (\ref{flat-sys}) we differentiate (\ref{flatcoord2}) with respect to $\l_j$ and use the expressions for Christoffel symbols in terms of the primary differential $\phi\,.$ 

A straightforward differentiation using (\ref{primary-variation}) shows that the functions listed in the theorem satisfy (\ref{flatcoord1}). To prove that  (\ref{flatcoord2}) holds for the functions $\{t^\iA\}$ we again consider the transformations $\{t^{\iA\delta}\}$ of these functions under the map $\l \to \l+\delta \;.$ Then, we find the action of the unit vector field on $\{t^\iA\}$ using the relation ${\bf e} (t^\iA) = (d/d\delta)t^{\iA\delta}\vert_{\delta=0}$  (see the proof of Theorem \ref{thm-primary}). 
$\Box$

The constant in (\ref{flatcoord2}) can be found by the method described in the proof of Theorem \ref{thm_flatcoord}; it is nonzero (equals $-1$) only if $t$ is the flat coordinate of the same type as the primary differential $\phi$ which defines the metric ${\bf ds}_\phi^{\bf 2}\,.$ Therefore we have the following corollary which shows again that the unit vector field is covariantly constant ({\bf F2}).
\begin{corollary}
\label{unityprop}
The unit vector field ${\bf e}$ (\ref{e}) in the flat coordinates $\{t^\iA\}$ of the metric ${\bf ds}_\phi^{\bf 2}$ defined by the primary differential $\phi=\phi_{t^{A_0}}$ has the form: ${\bf e}=-\d_{t^{A_0}}\;.$ 
\end{corollary}

Let us denote by $t^{\scriptscriptstyle{1}}$ the flat coordinate $t^{\iA_0}$ of the metric defined by the primary differential $\phi_{t^{A_0}}$ so that ${\bf e}=-\d_{t^1}\;.$

For each primary differential $\phi$ it is convenient to consider a multivalued differential $\Psi_\phi$ defined by:
\beqn
\Psi_\phi(P):=\left( {\rm v.p.} \int_{\infty^0}^P \phi \right) d\l \;,
\label{mult}
\eeqn
where the principal value near $\infty^0$ is defined by omitting the divergent part as a function of the local parameter $z_0\,.$ This differential
\begin{itemize}
\item is singular at the points $\infty^i\,.$ The nonconstant coefficients in expansions near $\infty^i$ are given by the flat coordinates $\{t^\iA\}$ of the corresponding metric ${\bf ds}_\phi^{\bf 2}\,.$ 
For $i\neq 0$ we have
\beqn
\Psi_\phi(P) \!\! \underset{P\sim{\infty^i}}{=} \!\! {\mbox{singular part}} +\! \left(\! v^i(n_i+1) z_i^{-n_i-2} + \!\! \sum_{\alpha=1}^{n_i} t^{i;\alpha} z_i^{-\alpha-1} + w^iz_i^{-1} + {\cal O}(1) \!\right) \! dz_i \;.
\label{Psi_expansion}
\eeqn
The coordinates $t^{0;\a}$ appear similarly in expansion in a neighbourhood of $\infty^0\,.$
\item transforms as follows under analytic continuation along the cycles $\{a_k;b_k\}\;:$
\begin{align}
\Psi_\phi (P^{a_k})-\Psi_\phi(P) & = - 2 \pi i \sum_{n=1}^g (\q^{-1})_{kn}  r^n d\l  + \delta_{\phi,\phi_{s^k}} d\l
- 2 \pi i  (\q^{-1})_{kk} \delta_{\phi,\phi_{r^k}} d\l
\label{psi_atwist} \;, \\
\Psi_\phi(P^{b_k})-\Psi_\phi(P) & = 2 \pi i r^k d\l  +  \delta_{\phi,\phi_{r^k}} 2 \pi i d\l  \;.
\label{psi_btwist}
\end{align}
\item is such that the combinations from (\ref{primary-periods}) of its $a$- and $b$-periods are given by coordinates \nolinebreak $s^k\,:$
\beqn
\oint_{a_k}\Psi_\phi + \sum_{n=1}^g(\q^{-1})_{kn} \oint_{b_n} \Psi_\phi = s^k  \;.
\label{psi_periods}
\eeqn
\end{itemize}

For each $\phi$ the multivalued differential $\Psi_\phi$ generates the set of primary differentials according to the following theorem.
\begin{theorem}
\label{PsiDeriv}
Derivatives of the multivalued differential $\Psi_\phi$ (\ref{mult}) with respect to the flat coordinates $t^\iA$ from Theorem \ref{thm_flatcoord} of the metric ${\bf ds}_\phi^{\bf2}$ are given by the corresponding primary differentials:
\beqn
\frac{\d\Psi_\phi}{\d t^\iA}=\phi_{t^\iA}\;
\label{Psider}
\eeqn
(we notice the independence of this derivative of the choice of a primary differential $\phi$).
\end{theorem}

{\it Proof.} Consider the differential $\d_{t^\iA}\Psi_\phi\,.$ From formulas (\ref{Psi_expansion}) - (\ref{psi_periods}) we see that its pro\-per\-ties (expansions near the points $\infty^i,$ transformations along the cycles $\{ a_k;b_k\}$ and the normalization (\ref{primary-periods})) coincide with analogous properties of the primary differential $\phi_{t^\iA}\;.$  Thus, the differentials $\d_{t^\iA}\Psi_\phi$ and $\phi_{t^\iA}$ are equal by virtue of Proposition \ref{Prop-unique}. 
$\Box$ 
\begin{corollary}
\label{Cor-der}
The derivatives of canonical coordinates $\{\l_j\}$ with respect to the flat coordinates $\{t^\iA\}$ of the metric ${\bf ds}_\phi^{\bf 2}$ are given by 
\beqn
\frac{\d\l_j}{\d t^\iA} = - \frac{\phi_{t^\iA}(P_j)}{\phi(P_j)} \;.
\label{flatder}
\eeqn
\end{corollary}

{\it Proof.} We shall use the reciprocity identity $\d_\a(fdg)_{g=const}=-\d_\a(gdf)_{f=const}\;.$ It holds for two functions $f$ and $g$ which can be locally expressed as functions of each other and some parameters $\{ p_\a \}\,,$ i.e.  $f=f(g; p_1,\dots,p_n)$ and $g=g(f; p_1,\dots,p_n)\,;$ where $\d_\a$ stands for a derivative with respect to the  parameter $p_\a\,.$ The reciprocity identity can be proven by differentiation of the identity $f(g(f;p);p)\equiv f$ with respect to $p_\a\;,$ i.e. $\d_\a gdf/dg + \d_\a f =0\,.$ For $f(P)=\int_{\infty^0}^P\phi$ and $g(P)=\l(P)$ we have 
\beqn
(\d_{t^\iA}\int^P_{\infty^0}\phi)d\l=-( \d_{t^\iA} \l(P) ) \phi(P) \;.
\label{reciprocity}
\eeqn
Using (\ref{Psider}) and $\l^\prime(P_j)=0$ we evaluate (\ref{reciprocity}) at the critical points $P=P_j$ to obtain (\ref{flatder}).
$\Box$

\subsection{Prepotential of Frobenius structures}

To complete the construction of Frobenius manifolds we need to show that requirement {\bf F3} holds. This can be done by constructing a prepotential, i.e. a function $F$ of flat coordinates $\{t^\iA\}$ of the corresponding metric ${\bf ds}_{\phi}$ such that
\beqn
\frac{\d^3 F_\phi}{\d_{t^A}\d_{t^B}\d_{t^C}}={\bf c}(\d_{t^A},\d_{t^B},\d_{t^C})={\bf ds}_{\phi}^{\bf 2}(\d_{t^A}\cdot\d_{t^B},\d_{t^C})\;.
\label{prep-def}
\eeqn
First, we need to define a pairing of differentials. 
Let $\omega^{(1)}$ and $\omega^{(2)}$ be two differentials on the surface $\surf$ holomorphic outside of the points $\infty^0,\dots,\infty^m$ with the following behaviour at $\infty^i\,:$
\beqn
\omega^{(\a)} = \sum_{n=-n^{(\a)}}^\infty c_{n,i}^{(\a)} z_i^n dz_i + \frac{1}{n_i+1} d \left( \sum_{n>0} r_{n,i}^{(\a)} \l^n \log \l \right) \;, \;\; P \sim \infty^i \;,
\label{inf-exp}
\eeqn
where $n^{(\a)}\in\Z$ and $c_{n,i}^{(\a)}\;,\;r_{n,i}^{(\a)}$ are some coefficients;  $z_i=z_i(P)$ is a local parameter near $\infty^i.$
Denote also for $k=1,\dots,g$ the coefficients $A_k^{(\a)}$ to be
\beqn
A_k^{(\a)} := \oint_{a_k}\omega^{(\a)} + \sum_{n=1}^g (\q^{-1})_{kn} \oint_{b_n} \omega^{(\a)} \;.
\label{Ak}
\eeqn
Again for $k=1,\dots,g\;,$  denote the transformations of differentials under analytic continuation along the cycles $\{a_k;b_k\}$ of the Riemann surface by:
\begin{alignat}{3}
&dp_k^{(\a)}(\l(P)) & := \omega^{(\a)} (P^{a_k}) - \omega^{(\a)} (P) \;,\hsp &p_k^{(\a)}(\l) &= \sum_{s>0} p^{(\a)}_{sk} \l^s \;,
\label{a-twist}\\
& dq_k^{(\a)}(\l(P)) &:= \omega^{(\a)}(P^{b_k}) - \omega^{(\a)}(P) \;, \hsp &q_k^{(\a)}(\l) &= \sum_{s>0}q^{(\a)}_{sk} \l^s \;.
\label{b-twist}
\end{alignat}
Note that the coefficients defined by (\ref{inf-exp}) - (\ref{b-twist}) for the primary differentials
do not depend on coordinates $\{t^\iA\}$ in contrast to the analogous coefficients for the differential $\Psi_\phi\,.$
\begin{definition}
Let $\omega^{(\a)}$ and $\omega^{(\b)}$ be  differentials which do not have singularities other than those described by (\ref{inf-exp})-(\ref{b-twist}). The {\bf pairing $\F[\;,\;]$} of such differentials is defined by:
\begin{align}
\begin{split}
\F[\omega^{(\a)}\;,\;\omega^{(\b)}] := \sum_{i=0}^m \left(\sum_{n\geq 0}\frac{c^{(\a)}_{-n-2,i}}{n+1}c^{(\b)}_{n,i} + c_{-1,i}^{(\a)} \mathrm{v.p.} \int_{P_0}^{\infty^i} \omega^{(\b)} - \mathrm{v.p.} \int_{P_0}^{\infty^i} \sum_{n>0}r_{n,i}^{(\a)} \l^n \omega^{(\b)} \right)\\
 + \frac{1}{2\pi i} \sum_{k=1}^g \left( - \oint_{a_k} q_k^{(\a)}(\l) \omega^{(\b)} + \oint_{b_k} p_k^{(\a)}(\l) \omega^{(\b)} + A_k^{(\a)} \oint_{b_ k} \omega^{(\b)} \right)\;,
\label{pairing}
\end{split}
\end{align}
where $P_0$ is a point on the surface such that $\l(P_0)=0\,.$  
\end{definition}

Note that the pairing is defined so that the following holds:
\begin{eqnarray}
\F[\phi_{t^\iA},\omega^{(\beta)}] = \I_{t^\iA}[\omega^{(\beta)}] \qquad \mbox{and} \qquad  \F[\Psi_\phi,\phi_{t^\iA}]  = \I_{t^\iA}[\Psi_\phi] \;.
\label{big}
\end{eqnarray}
Here $\omega^{(\beta)}$ is any differential for which the pairing is defined.

The last relation can be checked by a straightforward computation using (\ref{oper-commut}) and Proposition \ref{Prop-unique}. Now it is easy to prove the next theorem.
\begin{theorem}
\label{thm-prepotential} Let us choose one of the primary differentials $\phi$ given by Theorem \ref{thm-primary} and build the multivalued differential $\Psi_\phi$ (\ref{mult}). 
The following function gives a prepotential of the Frobenius structure
defined by the metric ${\bf ds}_\phi^{\bf 2}\;,$ multiplication (\ref{multiplication}) and the Euler field (\ref{Euler}) on the Hurwitz space $\covM_{g;n_0,\dots,n_m}$ outside the divisor $\Div$  (\ref{divisor}):
\beqn
 F_\phi = \frac{1}{2} \F [\Psi_\phi \;,\; \Psi_\phi ] \;.
\label{prepotential}
\eeqn
The second derivatives of the prepotential $F_\phi$ with respect to flat coordinates are given by the pairing of the corresponding primary differentials:
\beqn
\d_{t^A} \d_{t^B} F_\phi = \F[ \phi_{t^A}\;,\;\phi_{t^B}]\;.
\label{second-der}
\eeqn
\end{theorem}

{\it Proof.} Differentiating the function $F_\phi$ (\ref{prepotential}) using (\ref{big})  we obtain:
\beqn
\d_{t^\iA} F_\phi = \frac{1}{2} \F[\phi_{t^\iA},\Psi_\phi] + \frac{1}{2} \F[\Psi_\phi,\phi_{t^\iA}] = \F[\phi_{t^\iA},\Psi_\phi] \;.
\label{first-der}
\eeqn
Since the coefficients defined by (\ref{inf-exp}) - (\ref{b-twist}) for a primary differential do not depend on coordinates, the differentiation of both sides in (\ref{first-der}) gives
 (\ref{second-der}).
To find the third order derivatives of the function $F_\phi$ let us write the vector $\d_{t^{\iA}}$ (using  Corollary \ref{Cor-der}) in the form:
\beqn
\d_{t^\iA} = -\sum_{i=1}^L \frac{\phi_{t^\iA}(P_i)}{\phi(P_i)} \d_{\l_i} \;.
\label{flat_tangent}
\eeqn
A straightforward computation using (\ref{second-der}), (\ref{flat_tangent}) and the expression (\ref{primary-variation}) for derivatives of primary differentials with respect to canonical coordinates shows that the third derivatives coincide with the tensor ${\bf c}\,:$
\beqs
 \frac{\d^3 F_\phi (t)}{ \d_{t^A} \d_{t^B} \d_{t^C} } = {\bf c}(\d_{t^A},\d_{t^B},\d_{t^C}) =
 - \frac{1}{2} \sum_{i=1}^L \frac{\phi_{t^\iA}(P_i) \phi_{t^\iB}(P_i) \phi_{t^\iC}(P_i)}{\phi(P_i)} \;.
\eeqs
$\Box$

The prepotential $F_\phi$ satisfies the WDVV system (\ref{WDVV}) with respect to the flat coordinates $\{ t^\iA \}\,.$ Corollary \ref{unityprop} implies that $\d^3_{t^{\scriptscriptstyle{1}} t^\iA t^\iB} (F_{
\phi_{t^{\scriptscriptstyle{1}}}})\! =\!-{\bf ds}_{
\phi_{t^{\scriptscriptstyle{1}}}}^{\bf 2}(\d_{t^A},\d_{t^B})\;.$ Therefore, the matrix $F_1$ (\ref{matrices}) is constant since metric coefficients are constant in flat coordinates of the metric.

Let us denote by $\covM^{\phi,\q}$ the Frobenius structure on the Hurwitz space $\covM$ defined by the metric ${\bf ds}_{\phi}^{\bf 2}\,,$ where $\phi$ is one of the primary differentials from Theorem \ref{thm-primary}.
\begin{theorem}
Consider the flat metric ${\bf ds}_{\phi}^{\bf 2}$ given by (\ref{phi-metrics}). 
The nonvanishing matrix entries of this metric in the flat coordinates given by Theorem \ref{thm_flatcoord} are the following:
\beqs
{\bf ds}_\phi^{\bf 2} ( \d_{t^{i;\a}}, \d_{t^{j;\b}} )  = \frac{1}{n_i+1} \delta_{ij} \delta_{\a+\b,n_i+1} \;; \qquad
{\bf ds}_\phi^{\bf 2} ( \d_{v^i}, \d_{w^j} )  = \delta_{ij} \;; \qquad
{\bf ds}_\phi^{\bf 2} ( \d_{r^k}, \d_{s^l} )  = - \delta_{kl} \;.
\eeqs
\end{theorem}

{\it Proof.} The proof is given in \cite{Dubrovin}, p. 163 (see also \cite{doubles}); it uses the relation ${\bf ds}_\phi^{\bf 2} (\d_{t^\iA},\d_{t^\iB}) = {\bf e} \left( \F[\phi_{t^\iA},\phi_{t^\iB}] \right)$ and the representation of the unit vector field ${\bf e}$ via the action of the map $(\L,\l)\to(\L,\l+\delta)$ as in Proposition \ref{Wflat-prop}. 
$\Box$

The existence of the Euler vector field $E$ provides the quasihomogeneity (\ref{quasihomogeneity}) for the prepotential. Coefficients $\nu_\iA$ of quasihomogeneity coincide with those of the Frobenius structures in \cite{Dubrovin} (see also \cite{doubles}); these coefficients are  the coefficients of the Euler vector field written in the flat coordinates: $E = \sum_\iA \nu_\iA t^\iA \d_{t^\iA}\;.$ They can be found by computing the action of $E$ on the flat coordinates $\{t^\iA\}$ as in (\ref{temp1}). The charges $\nu$ of the constructed Frobenius manifolds can be computed from (\ref{temp1}); they are given by
$\nu = 1 - {2\a}/{(n_i+1)}$ for $\phi = \phi_{t^{i;\a}} \;;\;$
$\nu = -1$ for $\phi = \phi_{v^i}$ and $ \phi = \phi_{r^k}\;;\;$ 
$\nu = 1$ for $\phi = \phi_{\omega^i}$ and $\phi = \phi_{s^k}\;.$
A linear combination of the differentials corresponding to the same charge $\nu$ can be taken as a new primary differential for which a Frobenius structure can be built in the described way.

\section{Real doubles of the deformed Frobenius structures}
\label{SectDoubles}

Here we shall construct real doubles of the deformed semisimple Hurwitz Frobenius structures found in  Section \ref{SectDeform}. We use ideas of the work \cite{doubles} where the real doubles were found for the nondeformed Hurwitz Frobenius structures of \cite{Dubrovin}. The construction of \cite{doubles} is based on the properties of the Schiffer and Bergman kernels $\Omega(P,Q)$ and $B(P,Q)$ given by (\ref{Omegadef}), (\ref{Bdef}). Analogous structures for the deformations of Frobenius manifolds are obtained using the ``deformed" kernels $\Omega$ and $B\,,$ i.e. the bidifferentials $\Omt(P,Q)$ and $\Brt(P,\Qbar)$ (\ref{Omt-def}), (\ref{Brt-def}). Here we simply state main theorems; an essential part of the proofs can be found in \cite{doubles} and Section \ref{SectDeform}.  

In this section we consider the Hurwitz space $\covM_{g; n_0,\dots, n_m}$ as a space with local coordinates $\{ \l_1,\dots, \l_\iL; \lb_1, \dots, \lb_\iL \}. $ We shall denote it by $\covM_{g; n_0,\dots, n_m}^{\rm real}\;.$ The multiplication in the tangent space is again defined by $\d_{\l_i} \cdot \d_{\l_j} := \delta_{ij}\d_{\l_j}\;.$ The indices $i$ and $j$ range now in the set $\{ 1,\dots, L; \bar{1}, \dots, \bar{L}  \}$ and we define ${\l_{\bar{i}}} := {\lb_i}\,.$ This algebra obviously does not have nilpotents.
The Euler vector field has the standard form: $E := \sum_{i=1}^\iL \left( \l_i \d_{\l_i} + \lb_i \d_{\lb_i} \right) \;.$

Let us fix a point $P_0$ on the surface $\surf$ such that  $\l(P_0)=0\,,$ and let all basis cycles $\{a_k,b_k\}^g_{k=1}$ start at this point. Let us fix a constant symmetric nondegenerate matrix $\q$ such that the matrix $\Bt+\q$ is invertible (see (\ref{submanifold})). Denote by $f_\h$ and $f_\ah$ the holomorphic and antiholomorphic parts of a differential $f$ which can be represented in the form $f = f_\h + f_\ah\,.$ We say that a differential is of the $(1,0)$-type if in a local coordinate $z$ it has the form $f_\h=f_1(z)dz\;,$ and a differential is of the $(0,1)$-type
if in a local coordinate it has the form $f_\ah=f_2(\bar{z})d\bar{z}\,.$ 

Consider the following set of operations. Let $\tilde{\res}$ stand for the coefficient in front of $d\bar{z}/\bar{z}$ in the Laurent expansion of a differential. As before, $z_i$ is the local parameter in a neighbourhood of $\infty^{i}$ defined by $ z_i^{-n_i-1}(Q) = \l(Q) \;,\; Q \sim \infty^i \;.$\\

For $ i=0, \dots, m; \; \a=1, \dots, n_i $ we define:  
\begin{alignat*}{2}
 {\bf 1.} \;\; \I_{t^{i;\a}} [f(Q)] & := 
\frac{1}{\a} \; \underset{\infty^{i}}{\res}  \;  z_i^{-\a}(Q) f_{\h}(Q)  \qquad
& {\bf 2.} \;\; \I_{t^{\overline{i;\a}}} [f(Q)] & := 
 \frac{1}{\a} \; \underset{\infty^{i} }{\tilde{\res}} \; \bar{z}_i^{-\a}(Q) f_{\ah}(Q) \;.
\end{alignat*}

For $ i=1, \dots, m$ we define: 
\begin{alignat*}{2}
 {\bf 3.} \; \I_{v^i} [f(Q)]  &:= \; \underset{\infty^{i}}{\res} \; \l(Q) f_{\h}(Q) 
& {\bf 4.} \; \I_{v^{\bar{i}}} [f(Q)] & := \; 
\underset{\infty^{i}}{\tilde{\res}} \; \lb(Q) f_{\ah}(Q)  \\
{\bf 5.} \; \I_{w^i} [f(Q)] & := \mathrm{v.p.} \int_{\infty^0}^{\infty^i} f_{\h}(Q) & \qquad {\bf 6.} \; \I_{w^{\bar{i}}} [f(Q)] & := \mathrm{v.p.} \int_{\infty^0}^{\infty^i} f_{\ah}(Q) \;.
\end{alignat*}
As before, the principal value near infinity is defined by omitting the divergent part of an integral as a function of the corresponding local parameter. \\

For $k=1,\dots,g$ we define:
\begin{align*}
{\bf 7.} \;\; \I_{r^k} [f(Q)] & := - \oint_{a_k} \l(Q) f_{\h}(Q) - \oint_{a_k} \lb(Q) f_{\ah}(Q) - \sum_{j=1}^\iL (\q^{-1})_{kj} \oint_{b_j} \l(Q) f_\h(Q)  \\
{\bf 8.} \;\; \I_{u^k} [f(Q)] & := \oint_{b_k} \l(Q) f_{\h}(Q) + \oint_{b_k} \lb(Q) f_{\ah}(Q) \\
{\bf 9.} \;\; \I_{s^k} [f(Q)] & := \frac{1}{2\pi i} \oint_{b_k} f_\h(Q) \\
{\bf 10.} \;\; \I_{t^k} [f(Q)] & := - \frac{1}{2\pi i} \oint_{a_k} f_\h(Q) -\frac{1}{2 \pi i} \sum_{j=1}^\iL (\q^{-1})_{kj} \oint_{b_j} f_\h(Q) \;.
\end{align*}
Let us denote the set of operations by $\{ \I_{\xi^\iA} \},$ i.e. assume the index $\xi^\iA$ to belong to the set $\{ t^{i;\a}, t^{\overline{i;\a}};\; v^i, v^{\bar{i}}, w^i, w^{\bar{i}};\; r^k, u^k, s^k, t^k \}.$
Here we use $\xi^\iA$ as a formal index; later by $\xi^\iA$ we shall denote a flat coordinate on the Frobenius manifold.

The operations $\I_{\xi^\iA}$ define primary differentials $\Phi_{\xi^\iA}$ as follows.
\begin{alignat}{2}
\begin{split}
\Phi_{\xi^\iA}(P) &:= \I_{\xi^\iA} \left[ \Omt(P,Q) + \Brt(\Pbar,Q) \right]  \qquad \qquad \;\;\;\; {\mbox{for}} \qquad \xi^\iA \notin \{ r^k, u^k \}\;; \\
\Phi_{\xi^\iA}(P) &:= \I_{\xi^\iA} \left[ 2{\rm Re} \left\{  \Omt(P,Q) + \Brt(\Pbar,Q) \right\} \right] \qquad {\mbox{for}} \qquad \xi^\iA \in \{ r^k, u^k \}\;. 
\end{split}
\label{double-primary}
\end{alignat}
Variational formulas (\ref{q-variation}) imply the dependence of primary differentials (\ref{double-primary}) on canonical coordinates:
\begin{align}
\begin{split}
\frac{\d\Phi_{\xi^\iA}(P)}{\d\l_k} & = \frac{1}{2} \Phi_{\xi^\iA\h}(P_k) \left( \Omt(P,P_k) + \Brt(\Pbar,P_k) \right) \;; \\
\frac{\d\Phi_{\xi^\iA}(P)}{\d\lb_k} & = \frac{1}{2} \Phi_{\xi^\iA{\ah}}(P_k) \left( \Brt(P,\Pbar_k) + \overline{\Omt(P,P_k)} \right) \;.
\end{split}
\label{primary-dlambda}
\end{align}
Here $\Phi_{\xi^\iA\h}$ and $\Phi_{\xi^\iA\ah}$ are holomorphic and antiholomorphic parts of the differential $\Phi_{\xi^\iA}\,,$ respectively.  Relations (\ref{periods-rel}) for the periods of bidifferentials $\Omt$ and $\Brt$ imply similar relations on periods of differentials $\Phi_{\xi^\iA}$ ($\delta$ is the Kronecker symbol):
\beqn
\oint_{a_k} \Phi_{\xi^\iA} + \sum_{j=1}^g (\q^{-1})_{kj} \oint_{b_j}  \Phi_{\xi^\iA\h}  = \delta_{\xi^\iA, s^k}  \qquad {\mbox {and}} \qquad
 \oint_{b_k} \Phi_{\xi^\iA} = \delta_{\xi^\iA,t^k} \;.
\label{Phi-periods}
\eeqn
To prove relations (\ref{Phi-periods}) we integrate both sides of equalities (\ref{double-primary}) over $a$- and $b$-cycles. Due to the choice of the point $P_0,$ one can interchange integration and the operations $\I_{\xi^\iA}$ according to the rule (\ref{oper-commut}) (note that $\Wq(P,Q)$ and the sum $\Omt(P,Q) + \Brt(\Pbar,Q)$ have the same singularity structure).

The primary differentials $\{ \Phi_{\xi^\iA}\}$ (\ref{double-primary}) are alternatively specified as follows. They are differentials of the form $\Phi_{\xi^\iA}=\Phi_{\xi^\iA\h} + \Phi_{\xi^\iA\ah}$ which are normalized by relations (\ref{Phi-periods}) and possess the following properties (for proof see Theorem 2 of \cite{doubles}):
\begin{alignat*}{3}
& {\bf 1.}  \;\;\Phi_{t^{i;\a}}(P)  &&\sim (z_i^{-\a-1} + {\cal O}(1) ) dz_i + {\cal O}(1) d\bar{z}_i \;, \;\; P\sim\infty^i \;; \;\; \Phi_{t^{i;\a}} \;\; \mbox{is single valued on } \surf\,. \\
& {\bf 2.}  \;\; \Phi_{t^{\overline{i;\a}}}(P) &= \;\; &\overline{ \Phi_{t^{i;\a}}(P) } \;. \\         
& {\bf 3.} \;\;\Phi_{v^i}(P) \;  &&{\sim} -d\l + {\cal O}(1) \left( dz_i + d\bar{z}_i \right) \;, \;\; P\sim\infty^i \;; \;\; \Phi_{v^{\ib}} \;\;\mbox{is single valued on } \surf\,. \\
& {\bf 4\,.} \;\; \Phi_{v^{\ib}}(P) &= \;\; & \overline{\Phi_{v^i}(P)} \;. \\
&{\bf 5.}  \;\; \Phi_{w^i}(P): \;\; &&\underset{\infty^i}{\res} \; \Phi_{w^i} = 1 \;; \;\; \underset{\infty^0}{\res} \; \Phi_{w^i} = -1 \;. \;\; \Phi_{w^i} \;\; \mbox{is single valued on } \surf\,.\\
&{\bf 6.}  \;\; \Phi_{w^{\ib}}(P) &= \;\; & \overline{\Phi_{w^i}(P)} \;. \\
& {\bf 7.}  \;\; \Phi_{r^k}(P) :  && \mbox{has no poles}; &&\\
& \qquad \qquad \qquad  &&\Phi_{r^k}(P^{b_j}) - \Phi_{r^k}(P) = 2\pi i \, \delta_{kj}( d\l -d \lb) ;  \\
& \qquad \qquad \qquad &&\Phi_{r^k}(P^{a_j}) - \Phi_{r^k}(P) = -2 \pi i \, (\q^{-1})_{kj} d\l \;.  \\ 
& {\bf 8.} \;\; \Phi_{u^k}(P) : && \mbox{has no poles}; \\
& \qquad \qquad \qquad &&\Phi_{u^k}(P^{a_j}) - \Phi_{u^k}(P) = 2\pi i \, \delta_{kj}( d \l -  d \lb) \;.  \\
& {\bf 9.} \;\; \Phi_{s^k}(P) : && \mbox{single valued on $\surf\,$ and has no poles.}  \\
& {\bf 10.} \;\; \Phi_{t^k}(P) : && \mbox{single valued on $\surf\,$ and has no poles.} 
\end{alignat*}
Here, as before, $\l\!=\!\l(P),$ and  $z_i\!=\!z_i(P)$ is the local parameter at $P \!\sim\! \infty^i$ such that $\l \! = \! z_i^{-n_i - 1}\,.$
The indices $i\,,$ $k$ and $i;\a$ take values specified in the definition of operations $\I_{\xi^\iA}\;.$

The next theorem gives the Darboux-Egoroff metrics which satisfy requirements ${\bf F2}$ and ${\bf F4}$ (for proof see (\ref{temp1}), Theorem \ref{thm-primary}, and \cite{doubles} Propositions 7 and 9). 
\begin{theorem}
The metrics of the form
\beqn
{\bf ds}_\iPhi^{\bf 2} = \frac{1}{2} \sum_{i=1}^L \Phi_\h^2(P_i) (d\l_i)^2 + 
\frac{1}{2} \sum_{i=1}^L \Phi_{\ah}^2(P_i) (d\lb_i)^2 \;,
\label{Phimetric}
\eeqn
belong to the family (\ref{SBt-metrics1}), (\ref{SBt-metrics2}) of Darboux-Egoroff metrics.
Here $\Phi_\h$ and $\Phi_{\ah}$ are respectively the holomorphic and antiholomorphic parts of one of the primary differentials:  $\Phi(P)=\Phi_\h(P) + \Phi_{\ah}(P)\,.$  
The metric coefficients  satisfy $\;{\bf e}(\Phi_\h^2(P_i)) =0,$ $\;{\bf e}(\Phi_\ah^2(P_i)) =0$ and $E(\Phi_\h^2(P_i)) = -\nu \Phi_\h^2(P_i),$ $\;E(\Phi_\ah^2(P_i)) = -\nu \Phi_\ah^2(P_i)$ for some constant $\nu\,.$
\end{theorem}
A set of flat coordinates $\{ \xi^\iA \} := \{ t^{i;\a}, t^{\overline{i;\a}};\; v^i, v^{\bar{i}}, w^i, w^{\bar{i}};\; r^k, u^k, s^k, t^k \}$ (see Section \ref{FlatCoord}) of the metrics ${\bf ds}_{\iPhi}^{\bf2}$ (\ref{Phimetric}) is given by operations $\I_{\xi^\iA}$ applied to the primary differential $\Phi$ which defines the metric (see Theorem \ref{thm_flatcoord}, and \cite{doubles} Theorem 7). Namely, the flat coordinates of ${\bf ds}_{\iPhi}^{\bf2}$ are given by:
\vspace{0.3cm}

for $i=0,\dots,m;\;\a=1,\dots,n_i\;:\;$ $\;t^{i;\a} := -(n_i+1) \I_{_{t^{i;1+n_i-\a}}}[\Phi] ; \;\;\;
t^{\overline{i;\a}} := -(n_i+1) \I_{_{t^{\overline{i;1+n_i-\a}}}}[\Phi] ;
$

\vspace{0.2cm}
for $ i=1,\dots,m\;:\;\;$ $\;v^i := -\I_{w^i}[\Phi] \;; \;\;\;\;
v^{\bar{i}} := -\I_{w^{\bar{i}}}[\Phi] \;; \;\;\;\;
w^i := -\I_{v^i}[\Phi] \;; \;\;\;\;
w^{\bar{i}} := -\I_{v^{\bar{i}}}[\Phi] \;;
$ 

\vspace{0.3cm}
for $k=1,\dots,g\;:\;\;$ $r^k := \I_{s^k}[\Phi] \;; \;\;\;\; u^k := -\I_{t^k}[\Phi] \;; \;\;\;\;
s^k := \I_{r^k}[\Phi] \;; \;\;\;\; t^k := - \I_{u^k}[\Phi] \;.
$
\vspace{0.5cm}\\
As before, the unit vector field ${\bf e}$ is a vector field in the direction of the flat coordinate which has the same type as the differential defining the metric. Namely, in the flat coordinates of the metric ${\bf ds}_{\iPhi}^{\bf 2}$ with $\Phi=\Phi_{\xi^{\iA_0}}\,,$ the unit field is given by ${\bf e} = -\d_{\xi^{\iA_0}}\,.$ We shall denote this coordinate by $\xi^1$ so that ${\bf e} = -\d_{\xi^1}\,.$
\begin{lemma}
\label{lemmaJacobian} In the Hurwitz space outside the submanifold $\DivR$ defined by (\ref{submanifold}), the derivatives of canonical coordinates $\{\l_i\;;\lb_i\}$ with respect to flat coordinates $\{\xi^\iA\}$ of the metric ${\bf ds}_\iPhi^{\bf 2}$ are given by 
\beqs
\frac{\d\l_i}{\d\xi^\iA} = - \frac{\Phi_{\xi^\iA\h}(P_i)}{\Phi_\h (P_i)} \; , \qquad\qquad
\frac{\d\lb_i}{\d\xi^\iA} = - \frac{\Phi_{\xi^\iA\ah}(P_i)}{\Phi_\ah(P_i)} \; ,
\eeqs
where $\Phi(P)$ is the primary differential which defines the metric.
\end{lemma} 
The proof of this lemma repeats the proof of Lemma 4 in \cite{doubles}.

The analog of the multivalued differential (\ref{mult}) in the construction of real doubles is 
\beqn
\Psi_\iPhi(P)=\left(\vp\int_{\infty^0}^P\Phi_\h\right)d\l+\left(\vp\int_{\infty^0}^P\Phi_\ah\right)d\lb\;.
\label{mult-doubles}
\eeqn
The multivalued differential $\Psi_\iPhi$ again generates the set of primary differentials $\Phi_{\xi^\iA}$ according to the relation  $\d_{\xi^\iA}\Psi_{\iPhi}=\Phi_{\xi^\iA}\;.$

A prepotential of the Frobenius structure can be found with the help of the pairing of differentials which we shall define now. 

Let $\omega^{(\a)}(P) \;, \; \a=1,2\dots$ be a differential on $\surf$ which can be written as a sum of holomorphic and antiholomorphic differentials,
$\omega^{(\a)} = \omega_{\h}^{(\a)} + \omega_{\ah}^{(\a)} \; ,$ 
which are analytic outside of infinities and have the following behaviour at $P\sim\infty^i$  ($z_i=z_i(P)$ is a local parameter at $P\sim\infty^i$ such that $z_i^{-n_i-1}=\l$ ):
\begin{align}
\begin{split}
\omega_{\h}^{(\a)}(P) & =  \sum _{n=-n^{(\a)}_1} ^\infty c_{n,i}^{(\a)} z_i ^ n dz_i + \frac{1}{n_i+1} d \left( \sum_{n>0} r_{n,i}^{(\a)} \l^n \log\l  \right)  \;, \\
\omega_{\ah}^{(\a)}(P) & = \sum_{n=-n^{(\a)}_2 }^\infty c_{\nb,i}^{(\a)} \zb_i^n d\zb_i + \frac{1}{n_i+1} d  \left( \sum_{n>0} r_{\nb,i}^{(\a)} \lb^n \log\lb \right) \;,
\label{coeff_def}
\end{split}
\end{align}
where $n_1^{(\a)}, n_2^{(\a)} \in \Z \;;$ and $c_{n,i}^{(\a)} \; , \; r_{n,i}^{(\a)} \; , \; c_{\bar{n},i}^{(\a)} \; , \; r_{\nb,i}^{(\a)}$ are some complex numbers.
Denote also for $k = 1, \dots, g$ the combinations of periods:
\beqn
 A_k^{(\a)} := \oint_{a_k}\omega^{(\a)} + \sum_{j=1}^g (\q^{-1})_{kj}  \oint_{b_j} \omega(\a)_\h   \;, \hsp B_k^{(\a)} := \oint_{b_k} \omega^{(\a)} \;;
\label{Ak-doubles}
\eeqn
and the transformations along basis cycles:
\begin{alignat}{3}
\begin{split}
dp_k^{(\a)}(\l(P)) &:= \omega_{\h}^{(\a)}(P^{a_k}) - \omega_{\h}^{(\a)}(P)  \;, \hsp p_k^{(\a)} (\l) = \sum_{s>0} p^{(\a)}_{sk} \l^s \;,\\
dp_{\kb}^{(\a)} (\bar{\l}(P)) &:= \omega_{\ah}^{(\a)}(P^{a_k}) - \omega_{\ah}^{(\a)}(P)  \;, \hsp 
p_{\kb}^{(\a)} (\bar{\l}) = \sum_{s>0} p^{(\a)}_{\bar{s}\kb}\bar{\l}^s \;,\\
dq_k^{(\a)}(\l(P)) &:= \omega_{\h}^{(\a)} (P^{b_k}) - \omega_{\h}^{(\a)}(P)  \;, \hsp q_k^{(\a)} (\l) = \sum_{s>0} q^{(\a)}_{sk} \l^s \;,\\
dq_{\kb}^{(\a)}(\bar{\l}(P)) &:= \omega_{\ah}^{(\a)} (P^{b_k}) - \omega_{\ah}^{(\a)} (P)  \;, \hsp q_{\kb}^{(\a)} (\bar{\l}) = \sum_{s>0} q^{(\a)}_{\bar{s}\kb} \bar{\l}^s \;.
\end{split}
\label{qk-doubles}
\end{alignat}
Note that if the differential $\omega^{(\a)}$ is one of the primary differentials $\Phi_{\xi^\iA}$ (\ref{double-primary}), the coefficients defined by (\ref{coeff_def}) - (\ref{qk-doubles}) do not depend on coordinates $\{ \xi^\iA \}$. 
\begin{definition}
For any two differentials $\omega^{(\a)}$ and $\omega^{(\b)}$ which can be represented as a sum of holomorphic and antiholomorphic differentials, $\omega^{(\a)}=\omega^{(\a)}_\h+\omega^{(\a)}_\ah\;,$ and have only singularities of the type (\ref{coeff_def}), (\ref{qk-doubles}), the {\bf pairing $\F[\;,\;]$} is defined by:
\begin{multline}
\label{pairing-doubles}
\F[ \omega^{(\a)}\;, \;\omega^{(\b)} ] := \sum_{i=0}^m \left( \sum_{n\geq 0} \frac{c^{(\a)}_{-n-2,i}}{n+1} c^{(\b)}_{n,i} + c_{-1,i}^{(\a)} \mathrm{v.p.} \int_{P_0}^{\infty^i} \omega_{\h}^{(\b)} - \mathrm{v.p.} \int_{P_0}^{\infty^i} \sum_{n>0} r_{n,i}^{(\a)} \l^n \omega_{\h}^{(\b)} \right. \\
 \hspace{2.5cm} + \left. \sum_{n\geq 0} \frac{c^{(\a)}_{-\overline{n-2},i}}{n+1} c^{(\b)}_{\nb,i} + c_{-\bar{1},i}^{(\a)} \mathrm{v.p.} \int_{P_0}^{\infty^i} \omega^{(\b)}_{\ah} - \mathrm{v.p.} \int_{P_0}^{\infty^i} \sum_{n>0} r_{\nb,i}^{(\a)} \lb^n \omega_{\ah}^{(\b)} \right) \\
  + \frac{1}{2\pi i} \sum_{k=1}^g \left( -\oint_{a_k} q_k^{(\a)}(\l) \omega_{\h}^{(\b)} + \oint_{a_k} q_{\kb}^{(\a)}(\bar{\l}) \omega_{\ah}^{(\b)} + \oint_{b_k}p_k^{(\a)}(\l) \omega_{\h}^{(\b)} \right. \\
\left. - \oint_{b_k} p_{\kb}^{(\a)}(\lb) \omega_{\ah}^{(\b)} + A_k^{(\a)} \oint_{b_k} \omega_{\h}^{(\b)} - B_k^{(\a)} \left( \oint_{a_k} \omega_{\h}^{(\b)} + \sum_{j=1}^g (\q^{-1})_{kj}\oint_{b_j} \omega^{(\beta)}_\h\right) \right) \;.
\end{multline}
As before, $P_0$ is a point on $\surf$ such that $\l(P_0)=0,$ and the cycles $\{a_k, b_k \}$ all pass through $P_0\,.$
\end{definition}
The next theorem gives a prepotential of the Frobenius manifold, i.e. a function of flat coordinates $\{ \xi^\iA \}$ which satisfies the WDVV system. 
\begin{theorem} 
\label{thm-prep-doubles}
For each primary differential $\Phi$ consider the differential $\Psi_\iPhi(P)$ (\ref{mult-doubles}), multivalued on the surface $\surf\,.$ Consider the Frobenius structure defined  by the metric ${\bf ds}_\iPhi^{\bf 2}$ (\ref{Phimetric}), multiplication law $\d_{\l_i} \cdot \d_{\l_j} = \delta_{ij}\d_{\l_j}\,;\;$  $i ,j \in \{ 1,\dots, L; \bar{1}, \dots, \bar{L}  \}\,,\;$  ${\l_{\bar{i}}} := {\lb_i}\;,$ and the Euler vector field $E = \sum_{i=1}^\iL \left( \l_i \d_{\l_i} + \lb_i \d_{\lb_i} \right) \;.$ This Frobenius structure is defined on the manifold $\covM_{g;n_0,\dots,n_m}^{\rm real}$ outside the submanifold $\DivR$ of codimension one given by the equation $\det\,(\Bt+\q)=0\;.$ The prepotential $F_\iPhi$ for this Frobenius manifold is given by the pairing (\ref{pairing-doubles}) of the differential $\Psi_\iPhi$ with itself:
\beqn
F_\iPhi=\frac{1}{2}\F [ \Psi_\iPhi\;,\; \Psi_\iPhi ]\;. 
\label{prepotential-doubles}
\eeqn
The second order derivatives of the prepotential are given by:
\beqn
\d_{\xi^A} \d_{\xi^B} F_\iPhi = \F [ \Phi_{\xi^A}\;,\;\Phi_{\xi^B} ] - \frac{1}{4 \pi i} \delta_{\xi^\iA,s^k}\delta_{\xi^\iB,t^k}   + \frac{1}{4 \pi i} \delta_{\xi^\iA,t^k}\delta_{\xi^\iB,s^k} \;.
\label{secondDer}
\eeqn
Two last terms in (\ref{secondDer}) do not vanish only for the primary differentials $\Phi_{s^k}$ and $\Phi_{t^k}$ when the pairing $\F$ is not commutative.
The third order derivatives coincide with the tensor ${\bf c}\,:$
$\d_{\xi^A} \d_{\xi^B} \d_{\xi^C} F_\iPhi = {\bf c}(\d_{\xi^A},\d_{\xi^B},\d_{\xi^C}) := {\bf ds}_\iPhi^{\bf 2} \left( \d_{\xi^A}\cdot\d_{\xi^B},\d_{\xi^C} \right)\,.$
\end{theorem}
The proof of this theorem is analogous to proofs of Theorem \ref{thm-prepotential} and \cite{doubles}, Theorem 11.

The quasihomogeneity factors $\{ \nu_\iA \}$ (\ref{quasihomogeneity}) for the constructed deformations of real doubles of Frobenius manifolds coincide with those for the undeformed real doubles (\cite{doubles}, Proposition 11). 

Let us denote the constructed deformations of real doubles of Frobenius structures by $\covM^{{\rm real}\,\iPhi\,,\,\q}_{g; n_0,\dots,n_m}\,.$

The charges $\nu$ (see Definition \ref{defFrob}) of the manifolds $\covM^{{\rm real}\,\iPhi\,,\,\q}_{g; n_0,\dots,n_m}$ are as follows: if one chooses $\Phi := \Phi_{t^{i;\a}}$  or $\Phi := \Phi_{t^{\overline{i;\a}}}$ the charge is $\;\nu = 1 - {2\a}/({n_i+1}) \;; \;\;$  for $\Phi := \Phi_{v^i},\;\; \Phi := \Phi_{v^\ib},\;\; \Phi := \Phi_{r^k}\;\; $ or $\Phi := \Phi_{u^k}$ the charge is $ \;\nu = -1 \;; \;\;$ for $\Phi := \Phi_{\omega^i}, \;\; \Phi := \Phi_{\omega^\ib}, \;\; \Phi := \Phi_{s^k}\;$ or $\Phi := \Phi_{t^k}$ the charge is $\;\nu = 1\;.$

\section{$G$-function of the deformed Frobenius manifolds}
\label{SectG}

The $G$-function is a solution to the Getzler system introduced in \cite{Getzler}. The system is defined on an arbitrary semisimple Frobenius manifold. It was shown in \cite{DubZhang} that the Getzler system has a unique quasihomogeneous solution and that this solution has the form:
\beqn
G = \log \frac{\tau_\iI}{J^{\scriptscriptstyle{1/24}}}  \;.
\label{G-funct}
\eeqn
Here $J$ is the Jacobian of transformation from canonical to the flat coordinates,
$J = \det \left( \d_{\l_i} t^\alpha \right);$
and $\tau_\iI$ is the isomonodromic tau-function of the $\dim$-dimensional Frobenius manifold defined by 
\beqn
\frac{\d \log \tau_\iI}{\d \l_i} = \frac{1}{2} \sum_{j \neq i, j = 1} ^\dim \b_{ij}^2 (\l_i - \l_j) 
\;, \qquad i = 1, \dots, \dim \;.
\label{tauiso}
\eeqn
The $G$-function (\ref{G-funct}) of Dubrovin's Hurwitz Frobenius manifolds  \cite{Dubrovin} for the space of two-fold genus one coverings was computed in \cite{DubZhang}. In \cite{KokKorB}  the $G$-function was computed for an arbitrary Hurwitz Frobenius manifold of \cite{Dubrovin}.
As it was proven in \cite{KokKorG}, the isomonodromic tau-function $\tau_\iI$ for Hurwitz Frobenius manifolds can be expressed in terms of the so-called Bergman tau-function  $\tau_\iW$ on Hurwitz spaces: $\tau_\iW = \tau_\iI^{-2}\,,$ where the Bergman tau-function is defined as follows. 
Denote by $S^\iW$ the following term in the asymptotics of the kernel $W(P,Q)$ (\ref{W-def}) near the diagonal $ P \sim Q \,:$
\beqs
W(P,Q) \underset{Q \sim P}{=} \left( \frac{1}{(x(P) - x(Q))^2} + S^\iW(x(P)) + o(1) \right) dx(P) dx(Q)
\eeqs
(the quantity $6S^\iW(x(P))$ is called the Bergman projective connection \cite{Fay92}). Choosing the local parameter to be $x_i(P)  = \sqrt{ \l - \l_i }\;,$ we denote by $S^\iW_i$ the value of $S^\iW$ at a ramification point $P_i\,:$
\beqn
S^\iW_i = S^\iW(x_i)\Big|_{x_i = 0} \;.
\label{SWi}
\eeqn
Since the singular part of the $W$-kernel in a neighbourhood of the point $P_i$ does not depend on coordinates $\{ \l_j \},$ the Rauch variational formulas (\ref{W-variation}) imply ${\d_{\l_j} S^\iW_i} =  W^2(P_i,P_j)/2 \;.$
The symmetry of this expression with respect to the indices $i$ and $j$ provides the compatibility for the system of differential equations which defines the tau-function $\tau_\iW\,:$
\beqn
\frac{\d \log \tau_\iW }{\d \l_i} = - \frac{1}{2} S^\iW_i \;, \qquad  i=1, \dots, \dim \; .
\label{tau-W}
\eeqn

The $G$-function of the deformed Hurwitz Frobenius manifolds can be computed analogously to the method of \cite{KokKorG}.
\begin{theorem}
The $G$-function (\ref{G-funct}) of the deformed Hurwitz Frobenius structures $\covM^{\phi,\q}$ is given by 
\beqn
G = - \frac{1}{2} \log \left\{ \tau_\iW \frac{\det (\B+\q)}{\det\q} \right\} - \frac{1}{24} \log \prod_{i=1}^L \phi(P_i) + {\rm const} \;,
\label{G-funct-q}
\eeqn
the Bergman tau-function $\tau_\iW$ on the Hurwitz space is given by formula (1.5) from the paper \cite{KokKorB}.
The $G$-function (\ref{G-funct-q}) is defined on the Hurwitz space $\covM$ outside the divisor $\Div$ given by the equation $\det\,(\B+\q)=0\;.$
\end{theorem}

The constant $ \frac{1}{2} \log \left\{ \det\q \right\}$
is added in the right hand side to normalize the $G$-function so that it coincides with the $G$-function of \cite{Dubrovin} as $\q$ tends to infinity in such a way that $\Wq$ tends to $W$ (the function $G$ is defined up to an additive constant).

{\it Proof.}
According to the general formula (\cite{DubZhang}, p.36) the Jacobian of a Frobenius manifold is up to a constant given by the product of square roots of all nonvanishing coefficients of the Darboux-Egoroff metric ${\bf ds}^{\bf 2}$. Therefore, the Jacobian $J$ for the Hurwitz Frobenius manifold $\covM^{\phi,\q}$ is given by 
$J = {2^{-\iL/{\scriptscriptstyle 2}}}  \!\left( \prod_{i=1}^L \phi(P_i) \!\right)$ .

To compute the isomonodromic tau-function (\ref{tauiso}) for deformed Hurwitz Frobenius manifolds $\covM^{\phi,\q}$  we introduce a deformed Bergman tau-function $\tau_\iWq\,.$ The analogous to $S^\iW$ coefficient $S^\iWq(x(P))$ in the expansion of $\Wq(P,Q)$ near $P\simeq Q \simeq P_i$ is given by $S^\iWq_i =  S^\iW_i - \linebreak 2 \pi i \sum_{k,l=1}^\iL (\B+\q)^{-1}_{kl} \omega_k(P_i)\omega_l(P_i)\,.$
As a corollary of the variational formulas (\ref{Wq-variation}), we have 
\beqn
{\d_{\l_j} S^\iWq_i} =  \frac{1}{2} \, \Wq^2(P_i,P_j) = 2 \, \beta^2_{ij} \;,
\label{temp}
\eeqn
which allows to consistently define the tau-function $\tau_\iWq$ as follows:
\beqn
\frac{\d \log \tau_\iWq}{\d\l_i} = - \, \frac{1}{2} S^\iWq_i \;, \qquad  i=1, \dots, \dim \; .
\label{tau-Wq}
\eeqn
As is easy to verify using the definitions (\ref{tau-W}) and (\ref{tau-Wq}) of $\tau_\iW$ and $\tau_\iWq$, the ``deformed" and ``undeformed" tau-functions are related as follows:
\beqs
\tau_\iWq = \tau_\iW \det(\B+\q) \;.
\eeqs
Indeed, differentiation of the logarithm of this expression with respect to a branch point gives: $\d_{\l_j} \log \{\tau_\iW\det(\B+\q)\} = -S_i^\iW/2 + {\rm tr}\{ (\B+\q)^{-1}\d_{\l_i}(\B+\q) \}\,.$ The matrix $\q$ is independent of the branch points; using the derivatives of the matrix $\B$ given by the Rauch variational formulas (\ref{Rauch}) we prove that $\d_{\l_j} \log \{\tau_\iW\det(\B+\q)\} = - S_i^{\iWq}/2\,.$

Now, let us prove that the isomonodromic tau-function $\tau_{\iI,\q}$ defined by (\ref{tauiso}) for the manifolds $\covM^{\phi,\q}$ is given by $\tau_{\iI,\q} = (\tau_\iWq)^{-1/2}\,.$
First, we use relations (\ref{temp}) to rewrite the definition (\ref{tauiso}) of $\tau_{\iI,\q}$ in terms of the quantities $S^\iWq_i\;.$ To complete the proof it remains to  use
the equations $\; {\bf e}(S^\iWq_i) \equiv \sum_{j=1}^\iL \d_{\l_j} S^\iWq_i = 0 \;\; $ and  $\;\;  E(S^\iWq_i) \equiv \sum_{j=1}^\iL \l_j \d_{\l_j} S^\iWq_i = - S^\iWq_i $ which can be proven 
analogously to the similar relations (\ref{relations}) for coefficients of metrics (\ref{phi-metrics}).
$\Box$

The following theorem gives an expression for the $G$-function of the manifolds $\covM^{{\rm real}\,\iPhi,\q}\;.$
\begin{theorem}
The $G$-function of the deformations $\covM^{{\rm real}\,\iPhi,\q}$ of real doubles of Hurwitz Frobenius manifolds has the form:
\beqn
G = - \frac{1}{2} \log \left\{  \; | \tau_\iW |^2 \det \left({\rm Im} \B (\Bt+\q) \q^{-1} \right) \right\} - \frac{1}{24} \log \left\{ \prod_{i=1}^L \Phi_\h(P_i) \Phi_\ah(P_i) \right\} + {\rm const} \;,
\label{Gdoubles-q}
\eeqn
where $\tau_\iW$ is given by formula (1.5) of \cite{KokKorB}. The $G$-function (\ref{Gdoubles-q}) is defined on the Hurwitz space $\covM^{\rm real}$ outside the submanifold given by the equation $\det\,(\Bt+\q)=0\;.$
\end{theorem}

The constant $ \frac{1}{2} \log \left\{ \det\q \right\}$ is added in the right hand side to make the $G$-function (\ref{Gdoubles-q}) coincide with the $G$-function of real doubles of \cite{doubles} in the limit when the construction of deformations reduces to that of \cite{doubles}.

{\it Proof.}
The $G$-function (\ref{Gdoubles-q}) can be computed analogously to the $G$-function (\ref{G-funct-q}) of the Frobenius manifolds $\covM^{\phi,\q}$  by proving (similarly to \cite{KokKorG}, see also \cite{doubles}) the following expression for the isomonodromic tau-function defined by (\ref{tauiso}): $\tau_{\iI,\q} = (\tau_{\iOmt})^{-1/2}\,.$ (Note that the dimension of the Frobenius manifold $\covM^{{\rm real}\,\iPhi,\q}$ is $2L\;.$) The function $\tau_\iOmt$ is another analogue of the Bergman tau-function on Hurwitz spaces; it is defined as follows. Denote by $S^{\iOmt}_i$ the analogous to $S^\iW_i$ coefficient in expansion of the bidifferential $\Omt(P,Q)$ when both arguments are in a neighbourhood of the ramification point $P_i\,.$   Then, the following differential equations define the function $\tau_\iOmt\,:$
\beqn
\frac{\d\log\tau_\iOmt}{\d\l_i} = -\frac{1}{2} S_i^\iOmt \;, \qquad \frac{\d\log\tau_\iOmt}{\d\lb_i} = -\frac{1}{2} \overline{S_i^\iOmt} \;.
\label{tau-Omt}
\eeqn
Using differentiation formulas (\ref{Rauch}) and (\ref{Bt-variation}) for the matrices $\B$ and $\Bt\,,$ respectively, we prove that 
\beqs
\tau_{\iOmt} = \vert \tau_\iW \vert^2 \det({\rm Im}\B) \det(\Bt+\q)\,.
\eeqs

According to the general formula (\cite{DubZhang}, p.36), the Jacobian $J$ for the manifolds $\covM^{{\rm real }\Phi,\q}$ has the form: $J=\!2^{-\iL}\!\prod_{i=1}^L \Phi_\h(P_i) \Phi_\ah(P_i)\;.$ 
Substitution of this expression and the expression for the isomonodromic tau-function $\tau_{\iI,\q} = \left( \vert \tau_\iW \vert^2 \det({\rm Im}\B) \det(\Bt+\q) \right)^{-1/2}$ into (\ref{G-funct}) proves the theorem.
$\Box$

\newpage 
\section{Examples in genus one}
\label{SectExamples}

The bidifferential $\Wq(P,Q)$ (\ref{Wq-def}) is only different from $W(P,Q)$ in genus $g \geq 1,$ therefore the deformations of Hurwitz Frobenius structures are constructed only in positive genera. 

Consider the simplest Hurwitz space of two-fold coverings of genus one. According to the Riemann-Hurwitz formula (\ref{RH}), such coverings have four ramification points. Let one of them be over the point at infinity and denote the remaining three by $P_1, P_2, P_3\,.$ These coverings can be defined as the pairs $(\L,\l(\zeta))$ where $\L$ is the torus $\L = \C/\{ 2w, 2w^\prime \},\;$ $w, w^\prime \in \C,$ and $\l: \L \to \C P^1$ is the function 
\beqs
\l(\z) = \wp (\z) +c \;,
\eeqs
$\wp$ is the Weierstrass elliptic function and $c$ is a constant with respect to $\z\,.$ The ratio $\mu = w^\prime/w$ is the period of the torus, it is the $b$-period of the unique normalized holomorphic differential $\omega(\z) = d\z/(2w)\;,$ i.e.  $\mu = \oint_b \omega(\z)\,.$ The pair $(\L,\l)$ depends on three parameters: $w, w^\prime$ and $c\,.$ The branch points $\l_1,\; \l_2, \;\l_3$ of the covering can be expressed in terms of these parameters.  The $\z$-coordinates of ramification points are solutions to the equation $\l^\prime(\z)=0\,.$ This equation has three solutions in the domain $\C/\{ 2w, 2w^\prime \}$ due to the following relation on the $\wp$-function:
\beqs
\left( \wp^\prime(z) \right)^2 = 4 \left( \wp(z) - \wp(w) \right)\left( \wp(z) - \wp(w^\prime) \right)\left( \wp(z) - \wp(w+w^\prime) \right) \;,
\eeqs
where $\wp(w)+\wp(w^\prime)+\wp(w+w^\prime) =0\;.$ Hence, the branch points of the covering are given by 
$\l_1 = \wp(w) + c;\;$ $\l_2 = \wp(w^\prime) + c;\;$ $\l_3 = \wp(w + w^\prime) + c \;.$
The local parameter in a neighbourhood of a ramification point $P_i$ is $x_i(P)=\sqrt{ \l(P) -\l_i}\,.$ The branch points $\l_1,\l_2,\l_3$ play the role of local coordinates on the space of pairs $(\L,\l)\;;$ they are canonical coordinates on Frobenius manifolds. 

\subsection{$3$-dimensional Frobenius manifold and Chazy equation}

Here we give explicit formulas for ingredients of the Frobenius structure $\covM^{\phi_s,\q}_{1;1}$ on the Hurwitz space $\covM_{1;1}$ outside the divisor $\Div$ defined by the equation $\mu=-\q$ for some nonzero constant $\q\in \C\;.$  The differential $\phi_s$ (see Theorem \ref{thm-primary}) is given by
\beqn
\phi_s(\z) = \frac{1}{2\pi i} \oint_b \Wq( \z, \tilde{\z} ) = \frac{\q}{\mu+\q} \omega(\z).
\eeqn
The set of flat coordinates from Theorem \ref{thm_flatcoord} of the metric ${\bf ds}_{\phi_s}^{\bf 2}$ (\ref{phi-metrics}) is formed by the following three functions: 
\begin{align}
\begin{split}
&t_1 := s = - \oint_a \l\phi_s = - \frac{1}{2w} \int_x^{x+2w} (\wp(\z)+c) d\z = 
-\frac{\pi i}{4w^2}\gamma - c - \frac{\pi i }{\mu + \q} \frac{1}{2 w}\;, \\
& t_2 := t^{\scriptscriptstyle{0;1}} = \; \underset{\z = 0} {\res} \; \frac{1}{\sqrt{\l}} \left( \int _{\infty^0}^P \phi_s \right) d \l =  \frac{\q}{\mu + \q} \frac{1}{w} \;, \\
& t_3 := r = \frac{1}{2\pi i} \oint_b \phi_s = \frac{1}{2\pi i} \frac{\q \,\mu}{\mu + \q }  \;.
\label{flatcoord-ex}
\end{split}
\end{align}
Here $\g$ is such that $\int_x^{x+2w} \wp(\z) d\z = \pi i \g/(2w)$ for any $x\in\C\;,$ i.e. 
\begin{equation}
\gamma(\mu) = \frac{1}{3\pi i} \frac{\theta_1^{\prime\prime\prime}(0;\mu)}{\theta_1^\prime(0;\mu)} \;.
\label{gamma-ex}
\end{equation}
The metric ${\bf ds}_{\phi_s}^{\bf 2}$ in coordinates (\ref{flatcoord-ex}) is constant:
$ {\bf ds}_{\phi_s}^{\bf 2} = (1/2)(dt_2)^2 - 2 dt_1 dt_3 \;. $
The prepotential (\ref{prepotential}) has the form 
\begin{equation}
F = -\frac{1}{4} t_1 t_2^2 + \frac{1}{2}t_1^2t_3 - \frac{\pi i}{32} t_2^4 \left( \frac{1}{(1 - 2\pi i t_3/\q)^2}\gamma\left( \frac{2\pi i t_3}{1 - 2\pi i t_3/\q}\right) + \frac{2}{\q(1 - 2\pi i t_3/\q)}  \right) \;.
\label{prepotential-ex}
\end{equation}
This is a quasihomogeneous function: it satisfies $F_{\phi_s} (\kappa t_1, \kappa^{1/2} t_2, \kappa^{0} t_3 ) = \kappa^{2} F_{\phi_s}(t_1, t_2, t_3)$ for any nonzero constant $\kappa\;.$ The Euler vector field (\ref{Euler}) in coordinates (\ref{flatcoord-ex}) has the form: $E = t_1\d_{t_1} + (1/2) t_2 \d_{t_2}\,.$

To compute the function $G$ (\ref{G-funct-q}) for the manifold $\covM_{1;1}^{\phi_s,\q}$ we use the following expression for the function $\tau_\iW$ on the space $\covM_{1;1}$ (see \cite{KokKorB}): $\tau_\iW = \eta^2(\mu) \left( 2w\right)^{-1/4} \left(  \prod_{i=1}^L \omega(P_i) \right)^{-1/12} $ where $\eta(\mu)$ is the Dedekind eta-function $\eta(\mu) = (\theta^\prime_1(0))^{1/3}\,.$ Then, we have for the $G$-function:
\beqs
G = - \log \left\{ \eta \left( \frac{2 \pi i t_3}{1 - 2 \pi i t_3/\q} \right) (t_2)^\frac{1}{8} \left( 2 \pi i t_3/\q -1 \right)^{-\frac{1}{2}} \right\} \;.
\eeqs

In \cite{Dubrovin} a relationship was established between the $3$-dimensional WDVV system and the Chazy equation
\beqn
f^{\prime\prime\prime} = 6 ff^{\prime\prime} - 9 f^{\prime\; 2} .
\label{Chazy-ex}
\eeqn
Namely, the function of the form
\beqn
F = -\frac{1}{4} t_1 t_2^2 + \frac{1}{2}t_1^2t_3 - \frac{\pi i}{32}t_2^4 f(2 \pi i t_3)
\label{FrobChazy}
\eeqn
satisfies the WDVV system iff the function $f$ is a solution to the Chazy equation. 
The function $\g$ (\ref{gamma-ex}) satisfies the Chazy equation, and the Frobenius manifold $\covM^\omega_{1;1}$ of \cite{Dubrovin}  has the prepotential (\ref{FrobChazy}) with $f=\g\,.$ We shall call the Frobenius manifold $\covM^\omega_{1;1}$ \cite{Dubrovin} the {\it Chazy Frobenius manifold}. 

The group $SL(2,\C)$ maps one solution $f(\mu)$ of the Chazy equation to another solution $\tilde{f}(\mu)$ as follows:
\beqn
\tilde{f}(\mu) = f \left( \frac{a\mu + b}{c\mu + d} \right) \frac{1}{(c\mu + d)^2} - \frac{2c}{c\mu + d} \;,
\qquad  
\left( \begin{matrix} a&b\\c&d \end{matrix}  \right) \in SL(2,\C) \;.
\label{Chazy-sol}
\eeqn
Therefore there exists a $3$-parametric family of Frobenius manifolds of the form (\ref{FrobChazy}): 
\beqn
F = -\frac{1}{4} t_1 t_2^2 + \frac{1}{2}t_1^2t_3 - \frac{\pi i}{32}t_2^4 \left( \g \left( \frac{a 2 \pi i t_3 + b}{c 2 \pi i t_3 + d} \right) \frac{1}{(c 2 \pi i t_3 + d)^2} - \frac{2c}{c 2 \pi i t_3 + d} \right) \;.
\label{3-family}
\eeqn
In the case of integer coefficients $a\,,b\,,c\,,d\;,$ (\ref{3-family}) coincides with (\ref{FrobChazy}) with $f=\g\,.$

The manifold $\covM^{\phi_s,\q}_{1;1}$ (\ref{prepotential-ex}) gives a realization of a one-parameter subfamily of manifolds (\ref{3-family}) for $a=1\,,\; b=0\,,\; c= -1/\q\,,\; d =1\,.$
Thus we call it the {\it deformed Chazy Frobenius manifold}.

\subsection{Relationship to isomonodromic deformations}
\label{SectIso}

It was shown in \cite{BabKor} that the functions 
\begin{align}
\begin{split}
\Omega_1 = -\frac{1}{\pi \theta_2^2 \theta_3^2 } \left( 2d_\mu \log \theta_4 + \frac{1}{\mu+\q} \right) \;,&\hspace{1.2cm}
\Omega_2 = -\frac{1}{\pi  \theta_3^2 \theta_4^2 } \left( 2d_\mu \log \theta_2 +\frac{1}{\mu+\q} \right) \;, \\
\Omega_3 =-&\frac{1}{\pi i \theta_2^2 \theta_4^2 } \left( 2d_\mu \log \theta_3 +\frac{1}{\mu+\q} \right) \;
\label{Omegas}
\end{split}
\end{align}
satisfy the system of equations 
\begin{align}
\begin{split}
\frac{d \Omega_1}{dx} = \frac{1}{x} \Omega_2 \Omega_3\;,\hspace{1.2cm} 
&\frac{d \Omega_2}{dx} = - \frac{1}{x-1} \Omega_1 \Omega_3 \;,\hspace{1.2cm}
\frac{d \Omega_3}{dx} = \frac{1}{x(x-1)} \Omega_1 \Omega_2 \;, \\
\vspace{0.5cm}\\
&\Omega_1^2 +  \Omega_2^2 + \Omega_3^2 = - {1}/{4} \;,
\label{top}
\end{split}
\end{align}
where 
\beqs
x = \frac{\l_3 - \l_1}{\l_2 - \l_1} \;.
\eeqs
The  correspondence of notation in \cite{BabKor} to the one we use here is as follows: $\Omega_1^{\cite{BabKor}} = \Omega_3,$ $\Omega_2^{\cite{BabKor}} = i\Omega_1,$ $\Omega_3^{\cite{BabKor}} = -i\Omega_2,$ $\l_1^{\cite{BabKor}}=\l_3,$ $\l_2^{\cite{BabKor}}=\l_2,$ $\l_3^{\cite{BabKor}}=\l_1$ and $i\mu^{\cite{BabKor}}=\mu\,.$ The one-parameter solutions (\ref{Omegas}) were obtained as a certain limit of the general two-parametric family of solutions of (\ref{top}) found in \cite{Hitchin, BabKor}.

For any solution $\{\Omega_1,\Omega_2,\Omega_3\}$ to the system (\ref{top})  the formulas 
\beqn
\beta_{12} = \frac{\Omega_3}{\l_1-\l_2} \;,\hspace{1.2cm}
\beta_{23} = \frac{\Omega_1}{\l_2-\l_3} \;,\hspace{1.2cm}
\beta_{13} = \frac{\Omega_2}{\l_3-\l_1} 
\label{Omega-rot}
\eeqn
give rotation coefficients of some metric on the space $\covM_{1;1}$ which corresponds to a locally defined Frobenius structure (\cite{Dubrovin}, Proposition 3.5). The above system (\ref{top}) implies the flatness of this metric (equations (\ref{flat1})-(\ref{flat2})) and the following relation on the rotation coefficients: 
\beqn
\sum_{k=1}^3\l_k\d_{\l_k}\beta_{ij} = -\beta_{ij} \;.
\label{rotEuler}
\eeqn
\begin{proposition} 
\label{prop-top}
The rotation coefficients of the deformations $\covM_{1;1}^{\phi,\q}$ of Frobenius structures $\covM_{1;1}^\phi$ coincide with the coefficients (\ref{Omega-rot}) built from the solutions $\Omega_i$ (\ref{Omegas}) to system (\ref{top}).
\end{proposition}

{\it Proof.}
The space $\covM_{1;1}$ is the space of coverings of $\C P^1$ which have four simple ramification points $P_1, \; P_2, \; P_3 $ and $\infty^0\,.$ 
The Frobenius structures $\covM^{\phi,\q}_{1;1}$ described in Section \ref{SectDeform} have rotation coefficients $\beta_{12}=\Wq(P_1,P_2)/2,$ $\beta_{13}=\Wq(P_1,P_3)/2$ and  $\beta_{23}=\Wq(P_2,P_3)/2\,.$ 
Let us choose the $a$-cycle to encircle points $P_1$ and $P_3,$ and the $b$-cycle to encircle $P_2$ and $P_3\,.$ Then we have 
\beqn
\int_{\infty^0}^{P_1} \omega = \frac{\mu}{2}\;,\hspace{1.2cm}
\int_{\infty^0}^{P_2} \omega = \frac{1}{2}\;,\hspace{1.2cm}
\int_{\infty^0}^{P_3} \omega = \frac{\mu}{2} + \frac{1}{2}\;,
\label{cover-3}
\eeqn
where $\omega$ is the normalized holomorphic differential 
$ \omega = d\l {\Big/}\left(4w \sqrt{ (\l-\l_1)(\l-\l_2)(\l-\l_3)} \right)\,.$
For the bidifferential $W(P,Q):=d_\iP d_\iQ \log\theta_1(\int^Q_P \omega)\;,$ using relations (\ref{cover-3}), we get 
\begin{align*}
W(P_1,P_2) = - \omega(P_1) \omega(P_2) \frac{\theta_3^{\prime\prime}}{\theta_3} \;,\hspace{1.2cm}
&W(P_1,P_3) = - \omega(P_1) \omega(P_3) \frac{\theta_2^{\prime\prime}}{\theta_2} \;,\\
W(P_2,P_3) = - &\omega(P_2) \omega(P_3) \frac{\theta_4^{\prime\prime}}{\theta_4} \;,
\end{align*}
where $\theta_1(z) = - \theta[1/2,1/2](z)$ and $\theta_2=\theta[1/2,0](0),\; \theta_3=\theta[0,0](0), \; \theta_4=\theta[0,1/2](0)$ are the standard theta-constants. Then, using the Thomae formulas \cite{Fay92}  
\beqs
\pi^2 \theta_2^4 = (2\omega)^2 (\l_3-\l_1) \;,\hspace{1.2cm}
\pi^2 \theta_4^4 = (2\omega)^2 (\l_2-\l_3)\;,\hspace{1.2cm}
\pi^2 \theta_3^4 = (2\omega)^2 (\l_2-\l_1)\;,
\eeqs
and the heat equation for theta functions, $ \d^2_{zz}\theta[p,q](z) = 4 \pi i \d_{\mu}\theta[p,q](z)\;,$
we find that the rotation coefficients $\beta_{ij} =\Wq(P_i,P_j)/2$ are given by (\ref{Omegas}), (\ref{Omega-rot}). 
$\Box$

The system (\ref{top}) arises in the context of isomonodromic deformations of the matrix differential equation 
\beqs
\frac{d\Psi}{\d\l} = \left( \frac{A^0}{\l} + \frac{A^1}{\l-1} + \frac{A^x}{\l-x} \right) \Psi,
\eeqs
where $A^0, A^1, A^x \in sl(2,\C)\;$ and $\; \Psi \in SL(2,\C)\;.$ A solution $\Psi$ to this system has regular singularities at the points $\l=0,\; \l=1,\; \l=x$ and $\l=\infty\,.$ { \it Monodromy matrices} $M_\gamma$ are defined for a closed path $\gamma:[0,1] \to \C\setminus \{ 0,1,x \}$ encircling a singularity by 
\beqs
\Psi(\g(1)) = \Psi(\g(0)) M_\gamma \;.
\eeqs
The {\it isomonodromy condition} is the requirement for monodromy matrices to remain constant as $x$ varies. This is equivalent to the Schlesinger system for the matrices $A\,:$
\beqn
\frac{d A^0}{dx} = \frac{[A^x,A^0]}{x} \;, \hspace{1.2cm}
\frac{d A^1}{dx} = \frac{[A^x,A^1]}{x-1} \;, \hspace{1.2cm}
\frac{d A^x}{dx} = - \frac{[A^x,A^0]}{x} - \frac{[A^x,A^1]}{x-1} \;.
\label{Schlezinger}
\eeqn
This system implies that the functions ${\rm tr} (A^0)^2,\;{\rm tr} (A^1)^2,\;{\rm tr}
(A^x)^2$ are constant. If we fix them to be all equal $1/8$ then the functions 
\beqs
\Omega_1^2 = -(\frac{1}{8} + {\rm tr} A^1A^x), \hspace{1.2cm} 
\Omega_2^2 = -(\frac{1}{8} + {\rm tr} A^0A^x), \hspace{1.2cm}
\Omega_3^2 = -(\frac{1}{8} + {\rm tr} A^0A^1) 
\eeqs
give a solution to the system (\ref{top}). The system (\ref{top}) is also equivalent to the Painlev\'e-VI equation with coefficients $(1/8,-1/8,1/8,3/8)\,,$ see \cite{Dubrovin}, Appendix E, and \cite{Hitchin, BabKor, KitKor}.

\subsection{Real double of deformed Chazy Frobenius manifold}

Let us fix an imaginary constant $\q$ and consider the real Hurwitz space $\covM^{\rm real}_{1;1}$ with coordinates $\{\l_k;\lb_k\}$ outside the subspace defined by $\mu^\iOmega = -\q\,,$ where $ \mu^\iOmega := {\mu \bar{\mu}}/({\bar{\mu}-\mu}) = \oint_b \oint_{b} \Omega(P,Q)\,.$
The construction of a real double $\covM_{1;1}^{{\rm real}\,\Phi_s,\q}$ of the deformed Chazy Frobenius manifold $\covM_{1;1}^{\phi_s,\q}$ is based on the 
primary differential $\Phi_s$ (see (\ref{double-primary})). The differential $\Phi_s$ on a genus one surface is given by
\beqn
\Phi_s(\z) = \frac{\q}{\mu^\iOmega + \q} \left( \frac{\bar{\mu}}{\bar{\mu} - \mu}\omega(\z) + \frac{\mu}{\mu - \bar{\mu}}\overline{\omega(\z)} \right)\;.
\eeqn
The set of flat coordinates $\{\xi^\iA\}$ of the corresponding metric ${\bf ds}_{\iPhi_s}^{\bf 2}$ (\ref{Phimetric}) is given by the following six functions: 
\begin{align*}
t_1  &:= s = - \frac{\q}{\mu^\iOmega + \q} {\rm Re} \left\{ \frac{\bar{\mu}}{\bar{\mu}-\mu} \int_x^{x+2\omega} \!\!\!\! (\wp(\z)+c) \frac{d\z}{w} \right\}- \frac{\q}{\mu^\iOmega + \q} \frac{\bar{\mu}}{\bar{\mu}-\mu} \oint_b (\wp(\z)+c) \;,\\
t_2 & := t^{\scriptscriptstyle{0;1}} = \frac{\q}{\mu^\iOmega + \q}\frac{\bar{\mu}}{\bar{\mu}-\mu}\frac{1}{w} \;, 
 \qquad t_3  := r = \frac{\q}{\mu^\iOmega + \q}\frac{\mu^\iOmega}{2\pi i} \;,\\
t_4  &:= t = - \frac{\q}{\mu^\iOmega + \q} {\rm Re} \left\{ \frac{\bar{\mu}}{\bar{\mu}-\mu} \int_x^{x+2\omega^\prime} \!\!\!\!\! (\wp(\z)+c) \frac{d\z}{w} \right\}\;,\\
    \qquad t_5  &:= t^{\overline{\scriptscriptstyle{0;1}}} = \bar{t}_2 \;,
    \qquad t_6  := u =\frac{\q}{\mu^\iOmega + \q} \frac{1}{2\pi i} \frac{\bar{\mu}}{\bar{\mu}-\mu} \;. 
\end{align*}
The metric ${\bf ds}_{\iPhi_s}^{\bf 2}$ in these coordinates is constant:
$ {\bf ds}_{\Phi_s}^{\bf 2}=(dt_2)^2/2+(dt_5)^2/2-2dt_1dt_3+2dt_4dt_6\;.$

The prepotential (\ref{prepotential-doubles}) has the form: 
\begin{align}
 \begin{split}
   F_{\Phi_s}  & = -\frac{1}{4} t_1 t_2^2 - \frac{1}{4} t_1 t_5^2 + \frac{1}{2} t_1^2 t_3 - \frac{1}{2} t_1 t_4 (2 t_6 - \frac{1}{2 \pi i} ) \\
   & + t_3^{-1} \left( \frac{1}{4} t_2^2 t_4 (t_6 - \frac{1}{2 \pi i} ) + \frac{1}{4}t_4 t_5^2 t_6 + \frac{1}{2} t_4^2 t_6 ( t_6 - \frac{1}{2 \pi i}) + \frac{1}{16} t_2^2 t_5^2  \right)  \\
   & + 
   \frac{1}{32} t_2^4 \left( - \frac{1}{4 \pi i(t_6-t_3/\q)^{2}} \; \gamma \left( \frac{t_3}{t_6-t_3/\q} \right) + t_3^{-1} - \frac{1}{2\pi i t_3(t_6-t_3/\q)} \right)  \\
   & +  
   \frac{1}{32} t_5^4 \left( -\frac{\pi i} {(2 \pi i t_6 - 1)^2} \; \gamma \left( \frac{2 \pi i t_3}{1 - 2 \pi i t_6}    \right) +  t_3^{-1} + t_3^{-1} (2 \pi i t_6 - 1)^{-1} \right).   
 \label{Fs}
 \end{split}
\end{align}

The prepotential $F_{\iPhi}$ is a quasihomogeneous function: for any nonzero constant $\kappa$ it satisfies
\beqs
F_{\phi_s} (\kappa t_1, \kappa^{1/2} t_2, \kappa^{0} t_3,\kappa t_4, \kappa^{1/2} t_5, \kappa^{0} t_6  ) = \kappa^{2} F_{\phi_s}(t_1, \dots, t_6) \;.
\eeqs
The Euler vector field (\ref{Euler}) in coordinates (\ref{flatcoord-ex}) is given by: $E= t_1\d_{t_1} +  t_2/2 \d_{t_2} + t_4\d_{t_4} +  t_5/2 \d_{t_5}\;.$

The $G$-function (\ref{Gdoubles-q}) up to an additive constant has the form:
\beqn
G =  - \log \left\{ \eta \left( \frac{t_3}{t_6-t_3/\q} \right) \eta \left( \frac{2 \pi i t_3}{1- 2 \pi i t_6} \right) \left( t_2 t_5 \right)^\frac{1}{8} \left( \frac{t_3 }{(t_6-t_3/\q)(2 \pi i t_6 - 1 )}  \right)^\frac{1}{2}  \right\} \;,
\label{G-doubles-ex}
\eeqn
where we used the following relation for the Dedekind function: $\overline{\eta(\mu)} = \eta(-\bar{\mu})\,.$ 

The $G$-function (\ref{G-doubles-ex}) and the prepotential (\ref{Fs}) coincide with the corresponding objects of the real double construction of \cite{doubles} in the limit $\q \to \infty\,.$

\section*{Open problems}

Proposition \ref{prop-top} of Section \ref{SectIso} shows that rotation coefficients of the flat metric of the simplest deformed Frobenius manifold $\covM_{1;1}^{\phi_s,\q}$ are given by  formulas (\ref{Omega-rot}) with $\{\Omega_1,\Omega_2,\Omega_3\}$ being a one-parameter family (\ref{Omegas}) of solutions to the system (\ref{top}).
The general solution to the system (\ref{top}) which was found in \cite{Hitchin, BabKor} depends on two parameters.  For this solution, formulas (\ref{Omega-rot}) define  rotation coefficients which also correspond \cite{Dubrovin} to a Frobenius structure. 
The natural question is to find those structures which give a two-parametric deformation of Dubrovin's Hurwitz Frobenius manifold in genus one. 
The second problem will be to possibly generalize such deformations to Hurwitz spaces in arbitrary genus and find  real doubles of obtained structures.

Present work provides an indication that the construction of ``real doubles" of Dubrovin's Hurwitz Frobenius manifolds, proposed in \cite{doubles}, might have a universal character. To find a natural real double construction for an arbitrary Frobenius manifold and to clarify its meaning in applications to quantum cohomologies and other areas where Frobenius manifolds play a significant role is an interesting direction for further study. 

\vskip0.7cm
{\bf Acknowledgments.} I thank D. Korotkin for useful discussions.

\end{document}